\documentclass[longauth]{aa}

\usepackage{graphicx}
\usepackage{float}
\usepackage{txfonts}
\usepackage[dvipsnames,svgnames,x11names,hyperref]{xcolor}
\usepackage{lscape}
\newcommand{\angstrom}{\textup{\AA}}

\newcommand\T{\rule{0pt}{2.6ex}}       
\newcommand\B{\rule[-1.2ex]{0pt}{0pt}} 

\usepackage{natbib}
\bibpunct{(}{)}{;}{a}{}{,} 
\defcitealias{2020A&A...642A..46H}{H+2020}

\begin{document}

\title{The halo of M105 and its group environment as traced by planetary
nebulae populations\thanks{This research is based on data collected at Subaru Telescope, which is operated by the National Astronomical Observatory of Japan.
We are honoured and grateful for the opportunity of observing the Universe from Maunakea, which has the cultural, historical and natural significance in Hawaii. It is also based on observations made with the William Herschel Telescope operated on the island of La Palma by the Isaac Newton Group of Telescopes in the Spanish Observatorio del Roque de los Muchachos of the Instituto de Astrofísica de Canarias.}}

\subtitle{II. Using kinematics of single stars to unveil the presence of intragroup light around the Leo I galaxies NGC 3384 and M105}
\author{J. Hartke\inst{1,2} \and
       M. Arnaboldi\inst{2} \and
       O. Gerhard\inst{3} \and
       L. Coccato\inst{2} \and
       M. Merrifield\inst{4} \and
       K. Kuijken\inst{5} \and
       C. Pulsoni\inst{3} \and
       A. Agnello\inst{6} \and
       S. Bhattacharya\inst{7} \and
       C. Spiniello\inst{8,9}  \and
       A. Cortesi\inst{10} \and
       K. C. Freeman\inst{11} \and
       N. R. Napolitano\inst{12,13} \and
       A. J. Romanowsky\inst{14,15,16} 
       }

       \institute{European Southern Observatory,
              Alonso de Cordova 3107, Vitacura,
              Casilla 19001, Santiago de Chile, Chile\\
              \email{jhartke@eso.org}
              \and
              European Southern Observatory,
              Karl-Schwarzschild-Str. 2, 85748 Garching, Germany
              \and
              Max-Planck-Institut f\"{u}r Extraterrestrische Physik,
              Giessenbachstra{\ss}e, 85748 Garching, Germany
              \and
              School of Physics and Astronomy,
              University of Nottingham, NG7 2RD, United Kingdom
              \and
              Leiden Observatory, Leiden University,
              PO Box 9513, 2300~RA Leiden, The Netherlands
              \and
              DARK, Niels Bohr Institute, University of Copenhagen, Lyngbyvej 2, DK-2100 Copenhagen, Denmark
              \and
              Inter University Centre for Astronomy and Astrophysics, Ganeshkhind, Post Bag 4, Pune 411007, India
              \and
              Department of Physics,
              University of Oxford, Denys Wilkinson Building,
              Keble Road, Oxford OX1 3RH, United Kingdom
              \and
              INAF -- Osservatorio Astronomico di Capodimonte, Via Moiariello  16, 80131, Naples, Italy
              \and
              Departamento de Astronomia,
              Instituto de Astronomia, Geofisica e Ciencias Atmosfericas da USP,
              Cidade Universitaria,
              CEP:05508900 Sao Paulo, Brazil
              \and
              Research School of Astronomy \& Astrophysics Mount Stromlo
              Observatory, Cotter Road,
              2611 Canberra, Australia
              \and
              School of Physics and Astronomy, Sun Yat-sen University, DaXue Road 2, 519082 Zhuhai, China
              \and 
              CSST Science Center for the Guangdong-Hongkong-Macau Greater Bay Area, DaXue Road 2, 519082 Zhuhai, China
              \and 
              Department of Physics \& Astronomy, San José State University, One Washington Square, San Jose, CA 95192, USA    
              \and
              University of California Observatories, 1156 High St., Santa Cruz, CA 95064, USA
              \and
              Department of Astronomy and Astrophysics, University of California, Santa Cruz, CA 95064, USA
              }

          \date{\today}

       \abstract{M105 is an early-type galaxy in the nearby Leo~I group, the closest galaxy group to contain all galaxy types and therefore an excellent environment to explore the low-mass end of intra-group light (IGL) assembly.}
       {We present a new and extended kinematic survey of planetary nebulae (PNe) in M105 and the surrounding $30\arcmin\times30\arcmin$ in the Leo~I group with the Planetary Nebula Spectrograph (PN.S) to investigate kinematically distinct populations of PNe in the halo and surrounding IGL.}
       {We use PNe as kinematic tracers of the diffuse stellar light in the halo and IGL and employ photo-kinematic Gaussian mixture models to (i) separate contributions from  the companion galaxy NGC~3384, and (ii) associate PNe with structurally defined halo and IGL components around M105.} 
       {We present a catalogue of $314$ PNe in the surveyed area.  We firmly associate $93$ with the companion galaxy NGC~3384 and $169$ with M105. The PNe in M105 are further associated with its halo ($138$) and the surrounding exponential envelope ($31$). We construct smooth velocity and velocity dispersion fields and calculate projected rotation, velocity dispersion, and $\lambda_R$ profiles for the different components. PNe associated with the halo exhibit declining velocity dispersion and rotation profiles as a function of radius, while the velocity dispersion and rotation of the exponential envelope increase notably at large radii. The rotation axes of these different components are strongly misaligned.}
       {Based on the kinematic profiles, we identify three regimes with distinct kinematics that are also linked to distinct stellar population properties: (i) the \emph{rotating core} at the centre of the galaxy (within $1~R_\mathrm{eff}$) formed \emph{in situ} and dominated by metal-rich ([M/H]$\approx0$) stars and likely formed in situ, (ii) the \emph{halo} from $1$ to $7.5~R_\mathrm{eff}$ consisting of a mixture of intermediate-metallicity and metal-rich stars ([M/H]$>-1$), either formed \emph{in situ} or brought in through major mergers, and (iii) the \emph{exponential envelope} reaching beyond our farthest our farthest data point at 16 $R_\mathrm{eff}$, predominately composed of metal-poor ([M/H]$<-1$) stars. The high velocity dispersion and moderate rotation of the latter are consistent with that measured for the dwarf satellite galaxies in the Leo~I group, indicating that this exponential envelope traces the transition to the IGL.}

\maketitle
%

\section{Introduction}
The galaxy M105 (NGC~3379) in the Leo~I group figured predominantly in a lively debate on the dark matter content and its spatial distribution in massive early-type galaxies (ETGs) \citep{2003Sci...301.1696R,2005Natur.437..707D,2009mnras.395...76d,2009MNRAS.393..329N,2013MNRAS.431.3570M}. Using data obtained with the  Planetary Nebula Spectrograph (PN.S), \citet{2003Sci...301.1696R} and \citet{2007apj...664..257d} found that several intermediate luminosity elliptical galaxies -- M105 among them -- appeared to have low-mass and low-concentration DM halos, if any, based on their rapidly falling velocity dispersion profiles.
This is in contrast with inferences made on the massive dark halos of giant elliptical galaxies from multiple tracers, such as spectra of the integrated light \citep[e.g.][]{2000A&AS..144...53K, 2001AJ....121.1936G, 2006MNRAS.366.1126C}, X-ray profiles of the hot gas atmospheres \citep[e.g.][]{1994PASJ...46L..65A, 1999ASPC..163..153L, 2012ApJ...755..166H}, as well as from weak and strong gravitational lensing \citep[e.g.][]{2004ApJ...606...67H, 2006MNRAS.370.1008M, 2001ApJ...556..601W, 2006ApJ...649..599K, 2010ApJ...721L...1T, 2014MNRAS.445..162T}.
Follow-up studies discussed whether the apparent lack of dark matter in the halo of M105 was linked to the gravitational potential-orbital anisotropy degeneracy and viewing-angle effects \citep{2005Natur.437..707D, 2007apj...664..257d, 2009mnras.395...76d, 2009mnras.398..561w, 2013MNRAS.431.3570M}. Furthermore, the strong decrease of the line-of-sight (LOS) velocity dispersion may not continue in the outer halo; \citet{2018A&A...618A..94P} identified several ETGs where the decline of the LOS velocity dispersion was followed by an increase in the outer halo (e.g. NGC~1023, NGC~2974, NGC 4374, NGC~4472). Since \citet{2003Sci...301.1696R} measurements were confined within eight effective radii (i.e. within the field-of-view of a single PN.S pointing), they may have missed a kinematic transition at larger radii.

Using new, extended imaging and kinematic samples of PNe, we aim to investigate whether the LOS velocity dispersion profile decline in M105 continues in the outer halo or whether a change in kinematics is observed.
Our work builds on the first paper in this series, in which \citet[][hereafter \citetalias{2020A&A...642A..46H}]{2020A&A...642A..46H}, presented a wide-field photometric survey of PN candidates in M105 and the surrounding $0.5\times0.5\deg^2$. \citetalias{2020A&A...642A..46H} found a variation of the luminosity-specific PN number $\alpha$ with radius in the halo of M105, with the $\alpha$-parameter being seven times higher in the extended halo compared to the inner halo. 
They inferred that the PN population with a high $\alpha$-parameter is linked to a diffuse population of metal-poor stars \citep[$\mathrm{[M/H]} \leq -1.0$,][]{2016ApJ...822...70L}, whose light distribution is governed by an exponential surface brightness (SB) profile.
The light distribution in the inner halo is dominated by intermediate and metal-rich stars \citep[$\mathrm{[M/H]} > -1.0$,][]{2016ApJ...822...70L} following a S\'{e}rsic SB profile and a low-$\alpha$-parameter population of PNe.
In this paper, which is the second of the series, we wish to investigate whether the two PN populations that \citetalias{2020A&A...642A..46H} identified also possess distinct kinematic signatures and which constraints these would place on the assembly history of M105 and its group environment -- the Leo~I group. 

The Leo~I group is the closest group that contains all (i.e. early and late) galaxy types \citep{1975ApJ...202..610D}. The eleven brightest and most massive member galaxies can be further divided into two subgroups. Four are associated with the so-called Leo Triplet. The remaining seven, among them M105 and NGC~3384 that are the main subjects of this paper, are associated with the M96 (NGC~3368) group. In addition to these bright member galaxies, dwarf galaxies make up the largest number of group members: \citet{2018A&A...615A.105M} compiled the most recent catalogue of dwarf galaxies in the group, adding 36 new candidates to the previously known 52 dwarf galaxies.

With its low mass and proximity, the Leo~I group is an excellent environment to explore the low-mass end of intra-group light (IGL) assembly. Based on deep and wide-field photometry, \citet{2014apj...791...38w} attributed at most a few per cent of the light in their survey footprint to the IGL. Assuming that all PNe associated with the exponential SB profile trace the IGL, \citetalias{2020A&A...642A..46H} determine the fraction of PNe associated with the IGL to be 22\%, while the fraction in terms of stellar SB is 3.8\%. This low IGL fraction seemingly contradicts results from numerical simulations, which predict IGL fractions between 12\% and 45\% \citep{2006MNRAS.369..958S, 2006apj...648..936r}. However, it is unclear whether this mismatch is a resolution effect or whether a dimmer IGL is expected in lower-mass groups compared to more massive environments such as galaxy clusters\citep{2007ApJ...666...20P, 2014apj...791...38w}. Evaluating the dynamical status of the PNe populations at large radii is imperative for constraining the IGL properties in the Leo~I group.

This paper is organised as follows: in Sect.~\ref{sec:data}, we describe the data from photometric and slit-less spectroscopic surveys in the Leo~I group and the compilation of the final cross-matched data catalogues. Section~\ref{sec:decomposition} describes the decomposition of the sample into PNe associated with NGC~3384 and M105, based on photo-kinematic models. We describe the kinematics of PNe in the halo and envelope of M105 in Sect.~\ref{sec:kinematics}. We discuss our results in Sect.~\ref{sec:discussion} and put them into the context of the Leo~I group at large. We summarise and conclude our work in Sect.~\ref{sec:summary}. 

In this paper, we adopt a physical distance of 10.23~Mpc to M105. The corresponding physical scale is $49.6\;\mathrm{pc}/\arcsec$. This tip of the red giant branch (TRGB) distance was independently determined by \citet{2007AJ....134...43H,2007ApJ...666..903H} and \citet{2016ApJ...822...70L} and agrees well with that determined from the bright cut-off of the planetary nebula luminosity function (PNLF) determined by \citetalias{2020A&A...642A..46H} as well as with that derived from from SB fluctuation measurements \citep[SBF;][]{2001ApJ...546..681T}. The effective radius of M105, derived from broad-band photometry, is $R_\mathrm{eff,M105} = 54\farcs8 \pm 3\farcs5$ \citep{1990aj.....99.1813c}, corresponding to a physical scale of 2.7 kpc.

%
\section{The data}
\label{sec:data}
\begin{figure*}[h]
  \includegraphics[width=8.8cm]{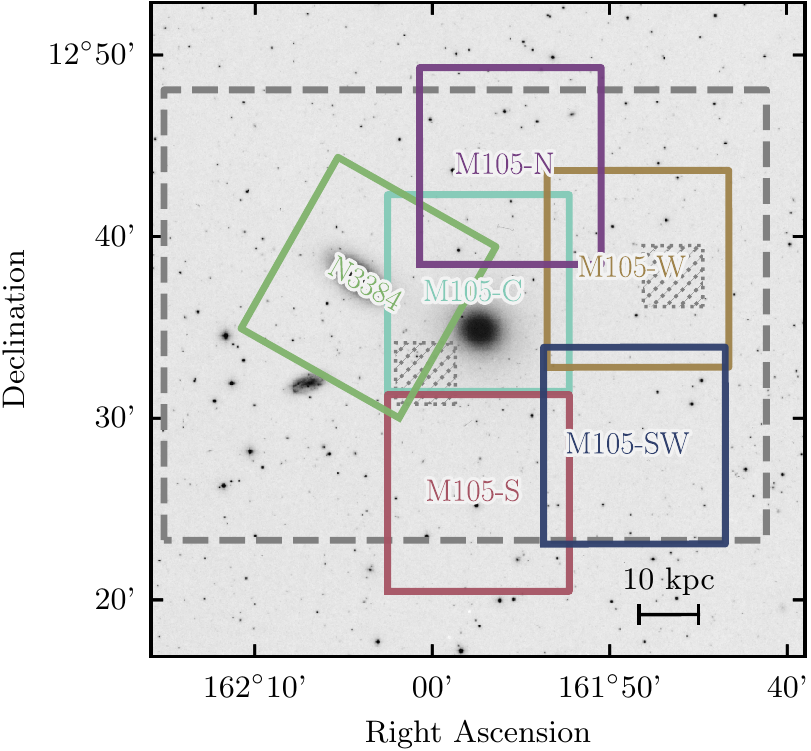}
  \includegraphics[width=8.8cm]{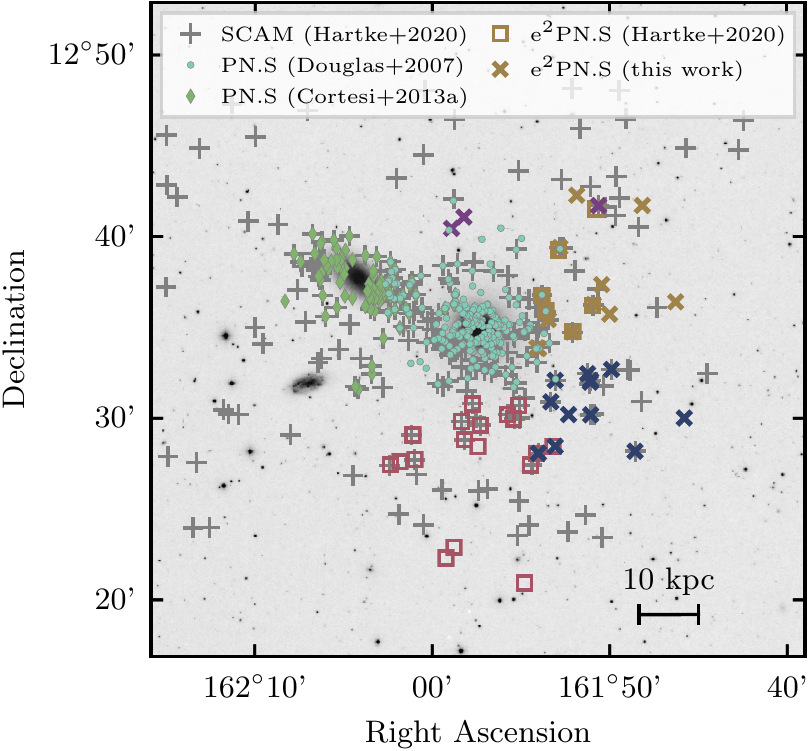}
  \caption{DSS-image of the Leo~I group with a scale-bar in the lower-right corner indicating a physical scale of 10 kpc. North is up, East to the left.
   \textit{Left:} Rectangles show the field outlines of the Surprime-Cam photometry survey (grey dashed outline), individual fields of the e$^2$PN.S survey (solid outlines), and the \textit{Hubble Space Telescope (HST)} fields analysed by \citet[][hashed rectangles]{2016ApJ...822...70L}.
    \textit{Right:} Overplotted are the PN candidates detected with Surprime-Cam \citep[][grey plus signs]{2020A&A...642A..46H}. The data from the ePN.S survey \citep{2018A&A...618A..94P} for NGC~3384 \citep[first published by][]{2013a&a...549a.115c} and M105 \citep[first published by][]{2007apj...664..257d} are indicated by cyan and green points and diamonds respectively. Squares denoted the PNe from the e$^2$PN.S survey already presented in \citepalias{2020A&A...642A..46H} and bold crosses newly detected PNe from this work. The colours of the symbols correspond to that of the field outlines in the left panel.}
  \label{fig:survey}
\end{figure*}

\subsection{Photometric survey}
\label{ssec:photometry}
\citetalias{2020A&A...642A..46H} presented the Subaru Surprime-Cam survey for PNe candidates in the Leo~I group. Here, we only briefly summarise the survey objectives, data reduction, and PN candidate identification and validation. They identified PN candidates from the combined use of narrow- [\ion{O}{iii}] ($\lambda_\mathrm{c,on} = 5500\;\angstrom$) and broad-band $V$-band ($\lambda_\mathrm{c,off} = 5029\;\angstrom$) images. 
Using CMD-based automated detection techniques \citep{2002aj....123..760a, 2003aj....125..514a}, \citetalias{2020A&A...642A..46H} identified $226$ PNe candidates within a limiting magnitude of $m_{5007,\mathrm{lim}} = 28.1$. These candidates are denoted with grey crosses in the left panel of Fig.~\ref{fig:survey}.
The photometric survey thus covers $2.6$ magnitudes from the bright cut-off of the PNLF at $m^{\star}_{5007} = 25.5$. The unmasked survey area (excluding image artefacts and regions with a high background value) covered $0.2365\;\mathrm{deg}^2$ on the sky, which corresponds to $67.7$~kpc along M105's major axis. The survey also covers the halos of NGC~3384 and NGC~3389. The survey footprint is outlined by the grey dashed rectangle in the right panel of Fig.~\ref{fig:survey}. The IDs, coordinates, and magnitudes of the PNe candidates brighter than the limiting magnitude are presented in Table~\ref{tab:scam}.

\subsection{The extremely extended PN.S ETG (e$^{2}$PN.S) survey in M105 and NGC~3384}
\begin{table*}
  \centering
  \caption{Summary of the PN.S observations used in this paper. The top part of the table refers to new observations, while the bottom part summarises data from published catalogues.}
  \label{tab:obs}
  \begin{tabular}{lllllrl}
  \hline
  \hline
  Field name & RA & dec & \multicolumn{1}{l}{Exposure time} & Year & $N_\mathrm{PN}$\tablefootmark{(a)} & References \T \\
   & [hh:mm:ss] & [dd:mm:ss] & \multicolumn{1}{l}{[hours]}  & & & \B \\
  \hline
  M105-S & 10:47:49.59 & +12:25:53.8 & 3.5 & 2017 & 18 & \T \\
  M105-W & 10:47:13.59 & +12:38:13.8 & 6.0 & 2017, 2019 & 13 &  \\
  M105-SW & 10:47:14.40 & +12:28:30.0 &  4.0 & 2019 & 11 &\\
  M105-N & 10:47:42.39 & +12:43:53.8 & 4.0 & 2019 & 2 & \\
  \hline
  M105-C & 10:47:49.60 & +12:36:54.0  & 18.8\tablefootmark{(b)} & 2002, 2003 &
217 & (1) \T \\
  N3384 & 10:48:14.40 & +12:37:12.0 & 5.95 & 2003-2011 & 94 & (2) \\
  \hline
  \hline
  \end{tabular}
  \tablebib{(1)~\citet{2007apj...664..257d}; (2)~\citet{	  2013a&a...549a.115c}}
  \tablefoot{
  \tablefoottext{a}{The number of PNe $N_\mathrm{PN}$ refers to the 3$\sigma$-clipped catalogue. The sum of $N_\mathrm{PN}$ presented in this table is larger than the total number of PNe in the main text as some PNe are observed in two fields.}
  \tablefoottext{b}{This exposure time is considerably longer compared to the other fields, but were obtained under a variety of seeing conditions. Had they been taken in nominal 1\arcsec seeing conditions, the equivalent would be 7 hours \citep[see][for details]{2007apj...664..257d}.}
  }
\end{table*}

The extremely extended PN.S ETG (e$^{2}$PN.S) survey includes data from the ePN.S survey \citep{2017iaus..323..279a,2018A&A...618A..94P}, namely two fields centred on M105 and NGC~3384 respectively. The first PN.S observations of M105 were part of the PN.S ETG survey \citep{2007apj...664..257d}. The resulting $214$ PNe are indicated with cyan circles in the right panel of Fig.~\ref{fig:survey} and in the left panel, where the field is outlined in the same colour and denoted with M105-C.

In 2017, we observed two additional fields (M105-W and M105-S). They are indicated by the brown and red rectangles in the left panel of Fig.~\ref{fig:survey}. These data were used to independently validate the photometric sample in \citetalias{2020A&A...642A..46H}. In 2019, M105-W was observed again, along with the fields M105-SW (blue rectangle) and M105-N (purple rectangle). The total exposure times per field are given in Table~\ref{tab:obs}.
NGC~3384 was first observed as part of the PN.S survey of S0 galaxy kinematics \citep{2013a&a...549a.115c}. In total, 101 PNe were observed; they are denoted with green diamonds in the right panel of Fig.~\ref{fig:survey}. The availability of these data will enable an improved photo-kinematic decomposition of the PN sample (see Sect.~\ref{sec:decomposition}).

\subsubsection{Field-to-field variation and catalogue matching within the e$^2$PN.S survey}
\label{ssec:kincat}
\begin{figure}
  \includegraphics[width=8.8cm]{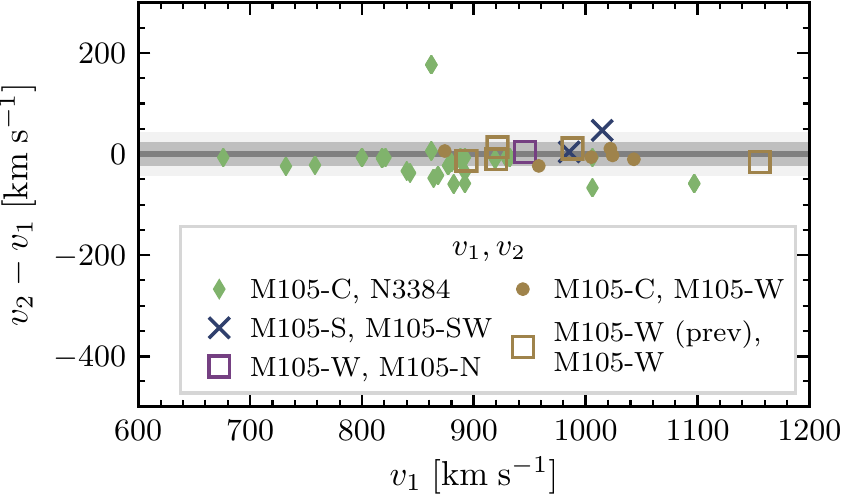}
  \caption{Comparison of the velocity measurements obtained for PNe that lie in the areas of intersection of two fields. The grey shaded regions indicate the nominal $1\sigma$ and $2\sigma$ PN.S velocity uncertainties of $\pm 20$ and $\pm 40\;\mathrm{km}\,\mathrm{s}^{-1}$, respectively. For the M105-W field, prev. indicates the previous measurements by \citetalias{2020A&A...642A..46H}.}
  \label{fig:vel-vel}
\end{figure}

The PN.S pipeline processes the survey data field-by-field. We, therefore, had to cross-match the individual catalogues to create a homogeneous master catalogue encompassing data from the six fields. We used an iterative coordinate-matching algorithm with a matching radius of $5\arcsec$. We chose this seemingly large radius due to the positional uncertainties on the PN.S data. As already stated in \citetalias{2020A&A...642A..46H}, we identify the $30$ PNe with measurements in the fields M105-C and N3384. In addition to that, we identify $2$ PNe in M105-S and M105-SW, $1$ PN in M105-N and M105-W, and $6$ in M105-C and M105-W.

Figure~\ref{fig:vel-vel} shows a field-to-field comparison of the velocities. The velocity measurements generally scatter about the one-to-one line, and the scatter is within the 2$\sigma$ PN.S velocity error (light grey shaded region). \citetalias{2020A&A...642A..46H} already discussed the nature of the 11 measurements that lie outside of the grey shaded region: these 11 PNe were observed close to one of the field edges in either of the two fields. For these objects, only the measurements taken closer to the respective field centre were included in the final catalogue. For the other PNe with repeat measurements, we included the mean velocities and magnitudes instead. The IDs of PNe with repeated measurements were concatenated in the final catalogue.

We also identified $5$ PNe in the western field based on 2017 data alone \citepalias{2020A&A...642A..46H} and in the deeper stack, including the data from 2019. These are denoted with orange squares in Fig.~\ref{fig:vel-vel}. They all scatter about the one-to-one line, and the scatter is within the velocity error of the PN.S. We conclude that there is no velocity offset between the observations taken in 2017 and 2019.

\subsubsection{Photometric calibration and catalogue matching with Surprime-Cam photometry}
\begin{figure}
  \includegraphics[width=8.8cm]{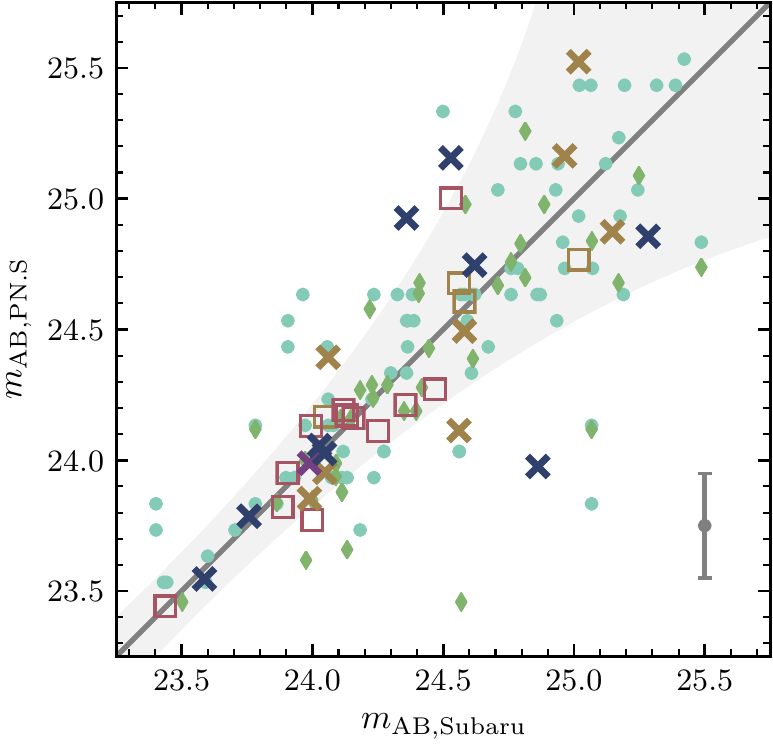}
  \caption{Comparison of the $AB$ magnitudes of PN candidates from \citetalias{2020A&A...642A..46H} with those from the e$^2$PN.S survey. The colour-coding is the same as in the left panel of Fig.~\ref{fig:survey}. The grey shaded region indicates where $99\%$ of Subaru sources would fall if one plotted two independent measurements of their magnitudes $m_\mathrm{AB,Subaru}$ against each other \citepalias{2020A&A...642A..46H}. The error bar in the lower-right corner denotes the typical magnitude uncertainty of the PN.S.}
  \label{fig:mag-mag}
\end{figure}

We matched the kinematic catalogue compiled in the previous section with the photometric one from \citetalias{2020A&A...642A..46H}. Like \citetalias{2020A&A...642A..46H}, we used a matching radius of $5\arcsec$ due to the relatively large positional uncertainty on the PN.S data. We used this matched sample to determine the zero-point magnitude of each of the newly-observed PN.S fields.  Fig.~\ref{fig:mag-mag} shows a comparison of the $AB$ magnitudes obtained with the PN.S after applying the zero-point offset with those from Surprime-Cam. The colour-coding is the same as in Fig.~\ref{fig:survey}.

The large scatter about the one-to-one line is not surprising, as the PN.S had not been optimised to obtain accurate photometry. While the photometric accuracy is acceptable in nearby galaxies \citep[e.g.][]{2006mnras.369..120m}, \citet{2018A&A...616A.123H} found a similarly large scatter of PN.S magnitudes compared to accurate Surprime-Cam ones in the Virgo galaxy M49. In order to convert the $AB$ magnitudes to $m_{5007}$, we used the relation determined by \citet{2003aj....125..514a}: $m_{5007} = m_{AB} + 2.49$.

\begin{figure}
  \includegraphics[width=8.8cm]{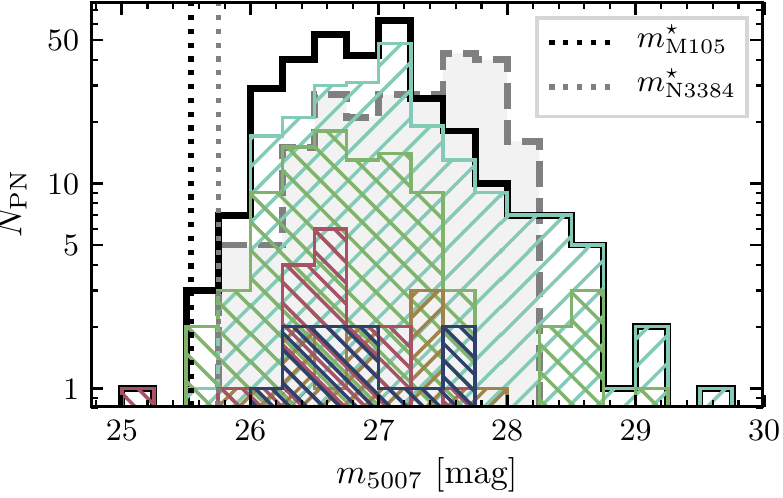}
  \caption{PNLF of all PNe from the e$^2$PN.S survey (black histogram) and its individual fields. The colour-coding of the histograms corresponds to that in Fig.~\ref{fig:survey}. The dashed gray histogram denotes the PNLF from the Surprime-Cam photometry. No completeness corrections have been applied to any of the LFs. The black and gray dotted vertical lines denote the PNLF bright cut offs corresponding to the distance of M105 and NGC~3384 respectively.}
  \label{fig:PNLF_fields}
\end{figure}

To gauge the depth of the e$^2$PN.S survey, we construct the PNLF of each of the fields. Figure~\ref{fig:PNLF_fields} shows the PNLF for each field with the same colour-coding as in Fig.~\ref{fig:survey}. The black histogram denotes the PNLF of all e$^2$PN.S PNe. For comparison, we also show the PNLF based on the Surprime-Cam photometry in grey. We only included data brighter than the limiting magnitude $m_{5007,\mathrm{lim}} = 28.1$.
The vertical dotted lines denote the PNLF bright cut-offs corresponding to the distances to M105 (black, $m^{\star}_{5007, \mathrm{M105}} = 25.54$) and NGC~3384 (grey, $m^{\star}_{5007, \mathrm{N3384}} = 25.76$) respectively. Except for one over-luminous object, all PNe are fainter than the bright cut-off $m^{\star}_{5007, \mathrm{M105}}$.

The deepest fields are centred on M105 and NGC~3384, with the faintest object nearly reaching 30th magnitude. However, the e$^2$PN.S survey is likely only complete to a limiting magnitude of $m_\mathrm{5007, lim} = 27.5$. This is the magnitude at which the number of PNe starts to decrease as a function of magnitude, contrary to the exponential increase that is theoretically expected (see also \citetalias{2020A&A...642A..46H}).

\subsubsection{Excluding velocity outliers}
\begin{figure*}
  \includegraphics[width=18cm]{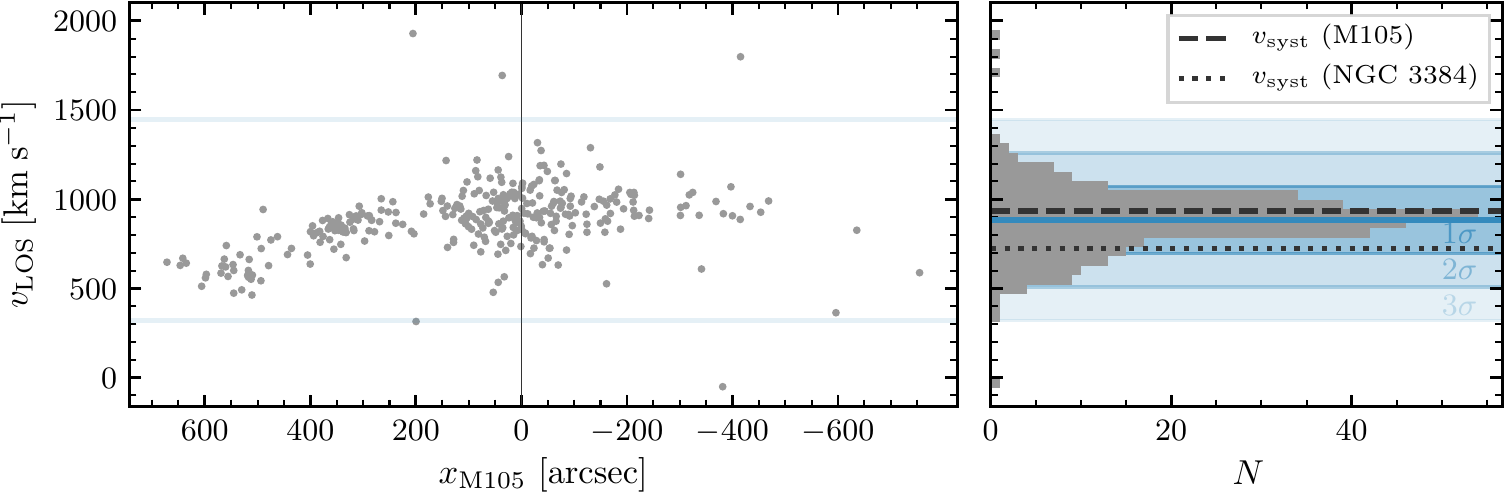}
  \caption{\textit{Left:} Phase-space of PN candidates in the e$^2$PN.S survey footprint, including PNe associated with M105 and NGC~3384. The major axis of M105 is denoted with $x_\mathrm{M105}$. \textit{Right:} Line-of-sight velocity histogram of the objects shown in the left panel. The dashed line denotes the systemic velocity of M105, and the solid blue line the mean velocity of the sample. The blue shaded regions denote the 1, 2, and $3\sigma$ intervals from the mean. Objects beyond $3\sigma$ of the mean velocity are likely redshifted background emission-line galaxies.}
  \label{fig:sigma_clip}
\end{figure*}

To remove velocity outliers that may be sources of contamination, we first calculated the mean velocity and velocity dispersion $\sigma$ of the whole sample -- including PNe associated with M105 and NGC~3384 -- using a robust fitting technique \citep{2010a&a...518a..44m}. The left panel of Fig.~\ref{fig:sigma_clip} shows the phase-space of PN candidates in the e$^2$PN.S footprint with the $x$-coordinate along the major axis of M105. The right panel of Fig.~\ref{fig:sigma_clip} shows the corresponding LOS velocity distribution (LOSVD) with the 1, 2, and $3\sigma$ intervals shaded in blue. The velocity range covered by the different filter configurations is $-400 < v \leq 2400\;\mathrm{km}\,\mathrm{s}^{-1}$.
We clipped four PNe with velocities outside the velocity range of $\pm 3\sigma$ about the robust mean.
Table~\ref{tab:obs} contains the final e$^2$PN.S catalogue after combining observations from multiple fields, correcting the magnitude zero-point, and $3\sigma$-clipping. In summary, the final catalogue, provided in Table~\ref{tab:pns_final}, contains coordinates, velocities, and magnitudes for $319$ PNe.

\subsubsection{Completeness and sources of contamination}
The completeness of the photometric catalogue has been extensively discussed in \citetalias{2020A&A...642A..46H}. Due to the nature of the PN.S instrument, we can immediately discard objects with a continuum, as the continuum appears as a stripe in slitless spectroscopy. This requirement excludes the majority of Ly-$\alpha$ emitters with a continuum redder than the [\ion{O}{iii}]5007\AA\ emission line. Background [\ion{O}{ii}] emitters at $z\simeq 0.34$ can be discarded as the spectral resolution of the PN.S is sufficient to resolve both emissions from the redshifted [\ion{O}{ii}] 3727 \AA\ doublet. However, we cannot use the second bluer line of the [\ion{O}{iii}] doublet commonly used to distinguish PNe from contaminants as the PN.S bandpass does not cover it. We therefore resorted to statistical methods to estimate the contamination from Ly-$\alpha$ galaxies without a measurable continuum, assuming that the number of background galaxies should be uniformly distributed in a velocity histogram \citep{2018mnras.477.1880s}.

Based on the presence of three objects in the velocity range $1400 < v \leq 2400\;\mathrm{km}\,\mathrm{s}^{-1}$ and one in the velocity range $-400 < v \leq 400\;\mathrm{km}\,\mathrm{s}^{-1}$ (see Fig.~\ref{fig:sigma_clip}), which are likely not PNe, we expect 7 objects such as background galaxies in the full velocity range, corresponding to a fraction of $2.2\%$. The result is very similar to that estimated by \citetalias{2020A&A...642A..46H} (2.6\%) based on a subset of the data and the estimate of \citet[][$\approx 2\%$]{2018mnras.477.1880s} for slitless spectroscopy of PNe the Fornax cluster.

\section{Galaxy membership assignment based on photo-kinematic models}
\label{sec:decomposition}

In order to decompose the kinematic PN sample into subcomponents associated with M105 and NGC~3384, we model the observed PN in position-velocity space using a luminosity-weighted multi-Gaussian model. The luminosity weights are pre-determined from broad-band photometry, while the kinematic parameters are free parameters of the Gaussian-mixture model. Gaussian-mixture models are widespread in the astronomical literature to disentangle multi-component LOSVDs \citep[see e.g.][]{2011ApJ...742...20W,2012MNRAS.419..184A,2013MNRAS.436.2598W,2014MNRAS.442.3299A,2018A&A...616A.123H,2018A&A...620A.111L}. In the following simple modelling we assume that there is no strong ongoing interaction between the two galaxies. This assumption is supported by the absence of strong tidal features or bridges of stars connecting the two galaxies \citep{2014apj...791...38w}.

\subsection{Disk-bulge decomposition of NGC~3384}
\label{ssec:disk-bulge}
To model the kinematics of NGC~3384, we use the disk-bulge decomposition method developed by \citet{2011MNRAS.414..642C} and subsequently applied to the PN.S survey of S0 galaxy kinematics \citep{2013MNRAS.432.1010C}, including the decomposition of NGC~3384 into its disk and bulge components.

The LOS velocity of the disk as function of azimuthal angle $\phi_\mathrm{3384}$ and radius $r_\mathrm{3384}$ measured with respect to the centre of NGC~3384 can be expressed as
\begin{equation}
    v_\mathrm{LOS}(r_\mathrm{3384}, \phi_\mathrm{3384}; v_\mathrm{rot}) = v_\mathrm{sys,3384} + v_\mathrm{rot}\sin(i)\cos(\phi_\mathrm{3384}),
\end{equation}
where $v_\mathrm{rot}$ is the galaxy’s mean rotation velocity and $i$ the inclination at which the disk is observed. The corresponding velocity dispersion is
\begin{equation}
\begin{split}
    \sigma_\mathrm{LOS}^2(r_\mathrm{3384}, \phi_\mathrm{3384}; v_\mathrm{rot}, \sigma_r, \sigma_{\phi}) = \sigma_r^2 \sin^2(i)\sin^2(\phi_\mathrm{3384}) + \\  + \sigma_{\phi_\mathrm{3384}}^2 \sin^2(i)\cos^2(\phi_\mathrm{3384}) + \sigma^2_z\cos^2(i),
\end{split}
\end{equation}
where $\sigma_r, \sigma_{\phi}$, and $\sigma_z$ are its components in cylindrical coordinates. In edge-on galaxies the contribution along the $z$-axis is small. Hence, in the following the term
$\sigma^2_z\cos^2(i)$ can be dropped. Assuming a Gaussian LOSVD, the disk kinematics can be described as
\begin{equation}
\begin{split}
    \mathcal{V}_\mathrm{N3384,disk}(v_k, r_{\mathrm{3384},k}, \phi_{\mathrm{3384},k}; v_\mathrm{rot}, \sigma_r, \sigma_{\phi}) =\\
    \frac{1}{\sqrt{2\pi}\sigma_\mathrm{LOS}} \exp\left( -\frac{(v_k - v_\mathrm{LOS})^2}{2\sigma_\mathrm{LOS}^2}\right)
\end{split}
\end{equation}

The contribution from the bulge is described as a Gaussian centred on the systemic velocity of NGC~3384 with velocity dispersion $\sigma_\mathrm{N3384,bulge}$, assuming it does not rotate significantly:
\begin{equation}
\begin{split}
    \mathcal{V}_\mathrm{bulge}(v_k, r_{\mathrm{3384},k}, \phi_{\mathrm{3384},k}; \sigma_\mathrm{N3384,bulge}) =\\
    \frac{1}{\sqrt{2\pi}\sigma_\mathrm{N3384,bulge}}\exp\left(\frac{(v_k - v_\mathrm{sys,3384})^2}{2\sigma_\mathrm{N3384,bulge}^2}\right).
\end{split}
\end{equation}

The two kinematic components of the LOSVD are weighted by the SB profiles as described in Sect.~\ref{ssec:photpriors}. In order to account for the variation of the LOSVD as a function of radius, we evaluate it in five concentric elliptical bins with the same geometry as the galaxy's isophotes centred on NGC~3384. Therefore the parameters $ v_\mathrm{rot}, \sigma_r, \sigma_{\phi}$, and $\sigma_\mathrm{N3384,bulge}$ are evaluated independently in each of the five bins. Following \citet{2011MNRAS.414..642C, 2013MNRAS.432.1010C}, we can reduce the parameter space by using the
epicycle approximation that holds for disk galaxies with approximately flat rotation curves \citep[see, e.g.][]{1987gady.book.....b} and links the radial velocity dispersion with the tangential
one: $\sigma_{\phi} = \sigma_r / \sqrt{2}$.

\subsection{Kinematic model of M105}
\label{ssec:kinematicM105}
\begin{figure*}
  \includegraphics[width=18cm]{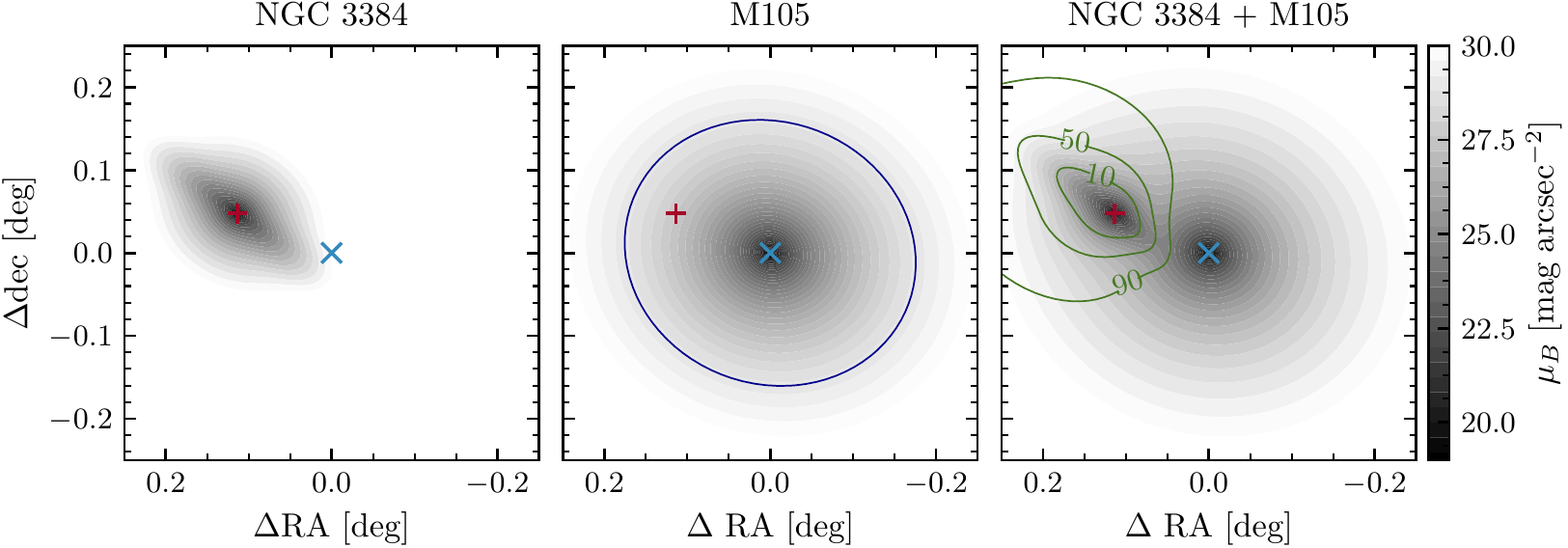}
  \caption{Modelled SB maps of NGC~3384 (left) and M105 (centre), with the centres of the galaxies marked with red and blue crosses, respectively. The right panel shows the composite SB map from both galaxies and green contours where M105 contributes $90\%$, $50\%$ and $10\%$ to the total light distribution.}
  \label{fig:SB}
\end{figure*}
Following the discovery of a PN population associated with an exponential SB profile in addition to the S\'{e}rsic halo \citepalias{2020A&A...642A..46H}, we model M105's kinematics LOSVD with two Gaussians; both centred on the galaxy's systemic velocity $v_\mathrm{sys,M105}$. They differ in their velocity dispersions, which we denote with $\sigma_\mathrm{M105,Ser}$ and $\sigma_\mathrm{M105,Exp}$, respectively:
\begin{equation}
\begin{split}
\mathcal{V}_\mathrm{M105,Ser}(v_k, r_{\mathrm{M105},k}; \sigma_\mathrm{M105,Ser}) = \\   \frac{1}{\sqrt{2\pi}\sigma_\mathrm{M105,Ser}} \exp\left( -\frac{(v_k - v_\mathrm{sys,M105})^2}{2\sigma_\mathrm{M105,Ser}^2}\right),
\end{split}
\end{equation}
\begin{equation}
\begin{split}
\mathcal{V}_\mathrm{M105,Exp}(v_k, r_{\mathrm{M105},k}; \sigma_\mathrm{M105,Exp}) = \\ \frac{1}{\sqrt{2\pi}\sigma_\mathrm{M105,Exp}} \exp\left( -\frac{(v_k - v_\mathrm{sys,M105})^2}{2\sigma_\mathrm{M105,Exp}^2}\right).
\end{split}
\end{equation}

Again, in order to account for any variation of the LOSVD with radius, the parameters $\sigma_\mathrm{M105,Ser}$ and $\sigma_\mathrm{M105,Exp}$ are evaluated in five concentric elliptical annuli with the same geometry as the galaxy's isophotes, but this time centred on M105. We assume that the LOS velocity dispersion of the exponentially distributed PN population is larger than that of the one following a S\'{e}rsic profile in each bin.

\subsection{Constraints on galaxy membership from broad-band photometry and RGB star number counts}
\label{ssec:photpriors}

\begin{figure*}
  \centering
  \includegraphics[width=8.8cm]{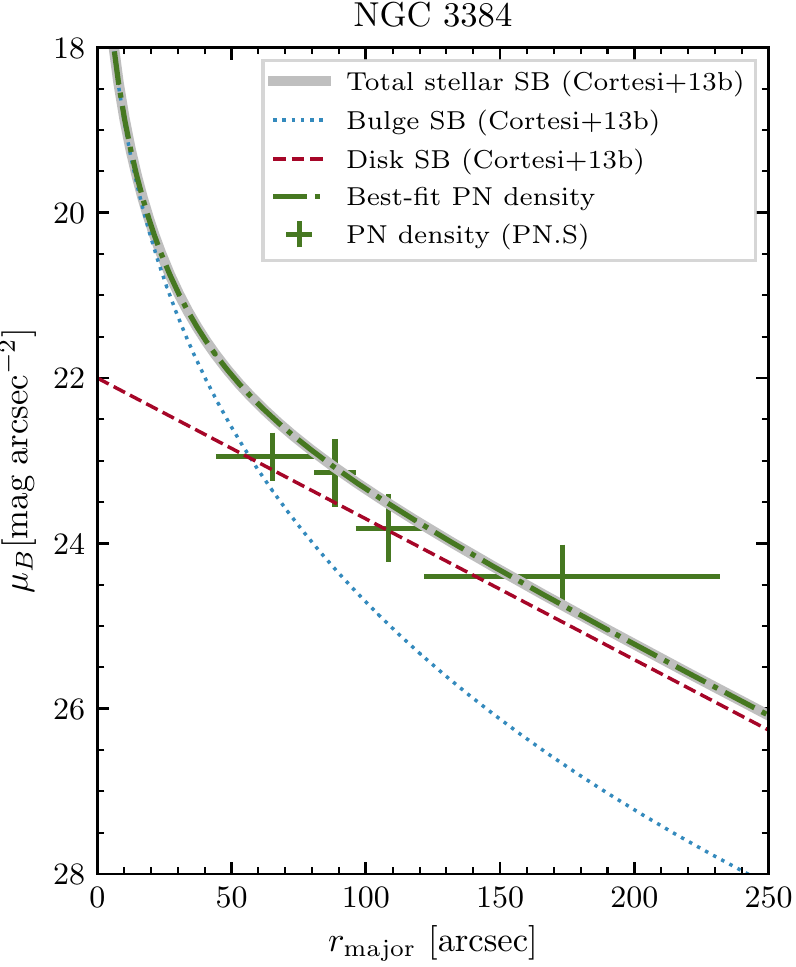}
  \includegraphics[width=8.8cm]{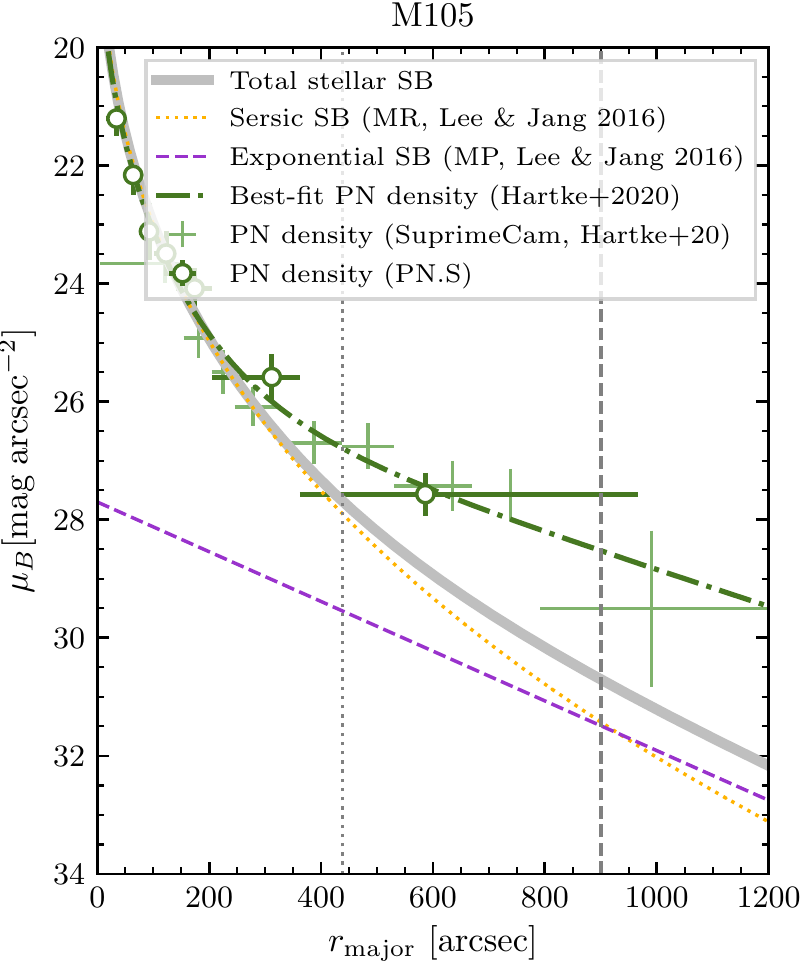}
  \caption{Stellar SB profiles of NGC~3384 and M105 in comparison with PN number density profiles scaled by the respective $\alpha$-parameters. In both panels, the solid grey line denotes the total stellar SB, and the dark green error bars the PN number density based on data from the PN.S.\textit{Left:} the stellar SB profile of NGC~3384 was decomposed into bulge (dotted red line) and disk (dashed blue line) by \citet{2013MNRAS.432.1010C} and the best-fit PN number density profile is denoted by the dash-dotted green line. \textit{Right:} the stellar SB profile of M105 was decomposed into S\'{e}rsic (dotted orange line) and exponential (dashed purple line) components by \citetalias{2020A&A...642A..46H} with priors on the number-densities of metal-poor (MP) and metal-rich (MR) RGB stars from \citet{2016ApJ...822...70L} and fit to the PN number density from Surprime-Cam data alone (light green error bars). The grey dotted vertical line denotes the radius at which the PN number density starts to deviate from the stellar SB profile and the dashed vertical line where the exponential SB starts to dominate the total light distribution.}
  \label{fig:PNdensity}
\end{figure*}

As the number density distribution of PNe is closely linked to the stellar light distribution, i.e. the galaxies' SB profiles \citep[see e.g.][and Fig.~\ref{fig:PNdensity}]{2009MNRAS.394.1249C}, we can obtain further constraints on which PN belongs to which galaxy component from broad-band photometry. The SB profiles of M105 and NGC~3384 are described by a combination of S\'{e}rsic profiles \citep{1963baaa....6...41s}. The intensity $I$ as a function of radius $R$ is:
\begin{equation}
  I_\mathrm{Ser}(R) = I_\mathrm{eff} \exp \left( -b_n \left[\left(\frac{R}{R_\mathrm{eff}}\right)^{1/n} -1\right]\right),
  \label{eqn:S\'{e}rsic}
\end{equation}
with
\begin{equation}
  b_n = 2 n - \frac{1}{3} + \frac{0.009876}{n}.
\end{equation}
The effective intensity $I_\mathrm{eff}$ is measured at the effective radius $R_\mathrm{eff}$ at which half the luminosity is enclosed, and the steepness of the profile is controlled by the S\'{e}rsic index $n$. The corresponding SB profile is
\begin{equation}
  \mu_\mathrm{Ser}(R) = \mu_\mathrm{eff} + \frac{2.5 b_n}{\ln10}\left[\left(\frac{R}{R_\mathrm{eff}}\right)^{1/n} -1\right],
\end{equation}
with the effective SB $\mu_\mathrm{eff} = -2.5\log I_\mathrm{eff}$.

An exponential SB profile with scale length $h$ is a special case of the S\'{e}rsic profile with index $n=1$:
\begin{equation}
  \mu_\mathrm{Exp}(R) = \mu_0 + 1.086\left(\frac{R}{h}\right)
\end{equation}
The scale length $h$ and central SB $\mu_0$ are related to the S\'{e}rsic quantities as follows:
\begin{align}
    R_\mathrm{eff} &= 1.678 \cdot h & \mu_\mathrm{eff} &= \mu_0 + 1.820
\end{align}
In the following, we describe the structural parameters of NGC~3384 and M105 that were derived in previous works. Table~\ref{tab:photometry} provides a summary in terms of S\'{e}rsic structural parameters, and Fig.~\ref{fig:SB} shows the resulting SB maps for the two galaxies separately, as well as a composite SB map. 

\begin{table*}
  \centering
  \caption{Structural parameters of M105 and NGC~3384 derived from broad-band surface photometry \citep{2014apj...791...38w, 2013MNRAS.432.1010C, 2006AJ....131.1163S} and PN number counts \citepalias{2020A&A...642A..46H} used as input for the photo-kinematic decomposition.}
  \begin{tabular}{llllllll}
    \hline
    \hline
    Component & $\mu_\mathrm{eff}$ & $R_\mathrm{eff}$ & Sersic $n$ & Inclination $i$ & Ellipticity $\epsilon$ & $P.A.$ & $\alpha_{2.5}$ \T \\
     & [mag arcsec$^{-2}$] & [\arcsec] & & [\degr] & & [\degr] & [$\times 10^8\,\mathrm{PN}\,L_\mathrm{bol}^{-1}$] \B\\
    \hline
    M105 halo & 21.73 & 57 & 2.75 & - & $0.111 \pm 0.005$ & $70.0\pm1.0$ & $1.00 \pm 0.11$ \T \\
    M105 exponential envelope  & 29.52 & 600 & 1 & - & $0.111 \pm 0.005$ & $70.0\pm1.0$ & $7.10 \pm 1.87$\\
    \hline 
    NGC 3384 bulge & 15.9 & 15.2 & 4 & - & 0.17 & 60.51 & $1.41\pm0.15$ \T \\
    NGC 3384 disk  & 20.07 & 107 & 1 & 70 & 0.66 & 52.50 & $1.41\pm0.15$ \\
    \hline 
    \hline
  \end{tabular}
  \label{tab:photometry}
\end{table*}

\subsubsection{NGC~3384}
\citet{2013MNRAS.432.1010C} analysed 2MASS $K$-band images \citep{2006AJ....131.1163S} with \textsc{Galfit} \citep{2010AJ....139.2097P} for the disk-bulge decomposition of NGC~3384. They found that the bulge's light distribution in the $K$-band is best described by a S\'{e}rsic profile
with effective SB $\mu_\mathrm{eff,bulge} = 15.9$, S\'{e}rsic index $n = 4$, effective radius $R_\mathrm{eff,bulge} = 15\farcs2$, ellipticity $\epsilon_\mathrm{bulge} = 0.17$, and position angle $PA_\mathrm{bulge} = 60\fdg51$.

The light distribution of the disk is described with an exponential profile with central SBs $\mu_\mathrm{0,disk} = 18.25$, scale length $h_\mathrm{N3384,disk} = 63\farcs73$, ellipticity $\epsilon_\mathrm{N3384,disk} = 0.66$, inclination $i = 70\degr$, and $PA = 52\fdg5$. \citetalias{2020A&A...642A..46H} determined a colour correction of $B-K = -3.75$ for the profiles derived from the 2MASS data by \citet{2013MNRAS.432.1010C}. The left panel of Fig.~\ref{fig:PNdensity} shows the total $B$-band SB profile of NGC~3384 (grey line) and the contributions from the galaxy's disk (blue dashed line) and bulge (red dotted line).

\subsubsection{M105}
The light distribution of M105 was first analytically described by \citet{1948AnAp...11..247D}, and is still regarded as one of the cornerstones of the \citeauthor{1948AnAp...11..247D} law today. Combining number density profiles of resolved red giant branch (RGB) stellar populations \citep{2016ApJ...822...70L} with deep wide-field imaging \citep{2014apj...791...38w}, and their wide-field photometric survey for PNe, \citetalias{2020A&A...642A..46H} established a two-component model for the SB profile of M105.
The light distribution in the inner halo is dominated by metal-rich stars following a S\'{e}rsic profile with $n_\mathrm{Ser}=2.75$, $R_\mathrm{eff,Ser} = 57\arcsec$ and $\mu_\mathrm{eff,Ser}=21.73$, denoted by the dotted orange line in the right panel of Fig.~\ref{fig:PNdensity}. In the outer halo, the fractional contribution from metal-poor stars, denoted by the dashed purple line, increases. Their light distribution is modelled with an exponential profile with $h_\mathrm{Exp} = 358\arcsec$ and $\mu_\mathrm{0,Exp} = 27.7$. \citet{2014apj...791...38w} derived constant a ellipticity $\epsilon_\mathrm{M105} = 0.111\pm0.005$ and a position angle of $PA = 70\fdg0 \pm 1\fdg0$ in the outer halo of M105.

\subsection{PN-specific frequency dependency on stellar population parameters}
\label{ssec:alphas}

Since the number of PNe has been shown to systematically vary with the colour of the parent stellar population \citep{2006MNRAS.368..877B,2015A&A...579A.135L, 2017A&A...603A.104H}, such dependency must be taken into account when using SB profiles derived from integrated starlight as weights for the kinematic model.
The connection between a PN population and the underlying stellar population is given by the PN-specific frequency, $\alpha$-parameter for short, that relates the total number of PNe $N_\mathrm{PN}$ to the total bolometric luminosity $L_\mathrm{bol}$ of the parent stellar population:
\begin{equation}
    N_\mathrm{PN} = \alpha L_\mathrm{bol}.
\end{equation}
For the bolometric correction, we use that derived from theoretical template galaxy models as function of the galaxies' colours \citep{2005mnras.361..725b, 2006MNRAS.368..877B}.
In the following, we will refer to $\alpha_{2.5}$ values, which are calculated in the magnitude interval  $m^* < m_{5007} \leq m^* + 2.5$. The observed PN number densities for NGC~3384 and M105 are indicated with green error bars in the left and right panels of Fig.~\ref{fig:PNdensity}, respectively. 

For NGC~3384, we assume that the $\alpha$-parameters for disk and bulge are equal (since, due to incompleteness in the centre of the galaxy we cannot calculate $\alpha_\mathrm{bulge}$ separately) and determined $\alpha_\mathrm{N3384} = (1.41 \pm 0.15) \times 10^8\,\mathrm{PN}\,L_\mathrm{bol}^{-1}$. We only considered PNe that lie within the green $50\%$ contour shown in the right panel of Fig.~\ref{fig:SB}. The resulting best-fit PN number density profile is indicated by the green dash-dotted line in the left panel of Fig.~\ref{fig:PNdensity}.

For M105, \citetalias{2020A&A...642A..46H} determined $\alpha$-parameters for PNe associated with the S\'{e}rsic and exponential SB profiles and found $\alpha_\mathrm{M105,Ser} = (1.00 \pm 0.11) \times 10^8\,\mathrm{PN}\,L_\mathrm{bol}^{-1}$ and $\alpha_\mathrm{M105,Exp} = (7.10 \pm 1.87) \times 10^8\,\mathrm{PN}\,L_\mathrm{bol}^{-1}$. The resulting best-fit PN number density profile is indicated by the green dash-dotted line in the right panel of Fig.~\ref{fig:PNdensity}.
Since the bulk of the light in M105 is contributed by metal-rich stars following the S\'{e}rsic profile, we use $\alpha_\mathrm{M105,Ser}$ as the denominator of the normalisation.
We therefore include the parameters $\tilde{\alpha}_{N3384} = \alpha_\mathrm{N3384}/\alpha_\mathrm{M105,Ser}$ and $\tilde{\alpha}_\mathrm{M105,Exp} = \alpha_\mathrm{M105,Exp}/\alpha_\mathrm{M105,Ser}$ that characterise the variation of $\alpha$ in the bulge and disk components of NGC~3384, as well as in the outer halo of M105 with respect to the $\alpha$-parameter of the main halo of M105.

\subsection{Bayesian likelihood}
The likelihood of a PN $k$ with observed position $RA, dec$ and velocity $v_k \pm \delta v_k$ can be expressed in terms of the velocity and surface-brightness distributions that we defined in the previous three sections. The model component weights are defined as follows:
\begin{subequations}
\begin{align}
    w_\mathrm{N3384,bulge} &= \frac{\tilde{\alpha}_\mathrm{N3384}\Sigma_\mathrm{N3384,bulge}}{n_\mathrm{tot}},\\
    w_\mathrm{N3384,disk} &= \frac{\tilde{\alpha}_\mathrm{N3384}\Sigma_\mathrm{N3384,disk}}{n_\mathrm{tot}},\\
    w_\mathrm{M105,Ser} &= \frac{\Sigma_\mathrm{M105,Ser}}{n_\mathrm{tot}},\\
    w_\mathrm{M105,Exp} &= \frac{\tilde{\alpha}_\mathrm{M105,Exp}\Sigma_\mathrm{M105,Exp}}{n_\mathrm{tot}},
\end{align}
\end{subequations}
where both the $\alpha$-parameters and the surface-brightness profiles $\Sigma$ are completely determined by the broad- and narrow-band photometry as described in Sects.~\ref{ssec:photpriors} and \ref{ssec:alphas}. The weights are normalised by 
\begin{equation}
\begin{split}
    n_\mathrm{tot} = \tilde{\alpha}_\mathrm{N3384}(\Sigma_\mathrm{N3384,bulge} + \Sigma_\mathrm{N3384,disk}) + \\
    + \Sigma_\mathrm{M105,Ser} + \tilde{\alpha}_\mathrm{M105,Exp}\Sigma_\mathrm{M105,Exp}
\end{split}
\end{equation}
The corresponding likelihood for a PN $k$ at its position in phase-space is
\begin{equation}
\begin{split}
    \mathcal{L}_k = w_\mathrm{N3384,bulge}\mathcal{V}_\mathrm{N3384,bulge} + w_\mathrm{N3384,disk}\mathcal{V}_\mathrm{N3384,disk} + \\
    + w_\mathrm{M105,Ser}\mathcal{V}_\mathrm{M105,Ser} + w_\mathrm{M105,Exp}\mathcal{V}_\mathrm{M105,Exp}.
\end{split}
\end{equation}
The total likelihood is the product of the individual likelihoods that is calculated for each \emph{individual} PN with coordinates ($RA, dec$) and LOS velocity $v \pm \delta v$:
\begin{equation}
    \mathcal{L} = \prod_{k = 0}^{N}\mathcal{L}_k.
\end{equation}
This method allows us to exploit the information of every PN without the explicit need for binning in velocity space. As discussed in Sects.~\ref{ssec:disk-bulge} and \ref{ssec:kinematicM105}, we evaluate the five LOSVD parameters $v_\mathrm{rot}, \sigma_r, \sigma_\mathrm{N3384,bulge}, \sigma_\mathrm{M105,Ser}, \sigma_\mathrm{M105,Exp}$ independently in five elliptical annuli, centred on the respective galaxy (solid and dashed ellipses in Fig.~\ref{fig:decomposition}). This results in a 25-dimensional parameter space that is explored using the ensemble-based MCMC sampler \textsc{emcee} \citep{2013PASP..125..306F}. 

\subsection{Priors and posteriors}
\begin{figure*}
    \centering
    \includegraphics[width=18cm]{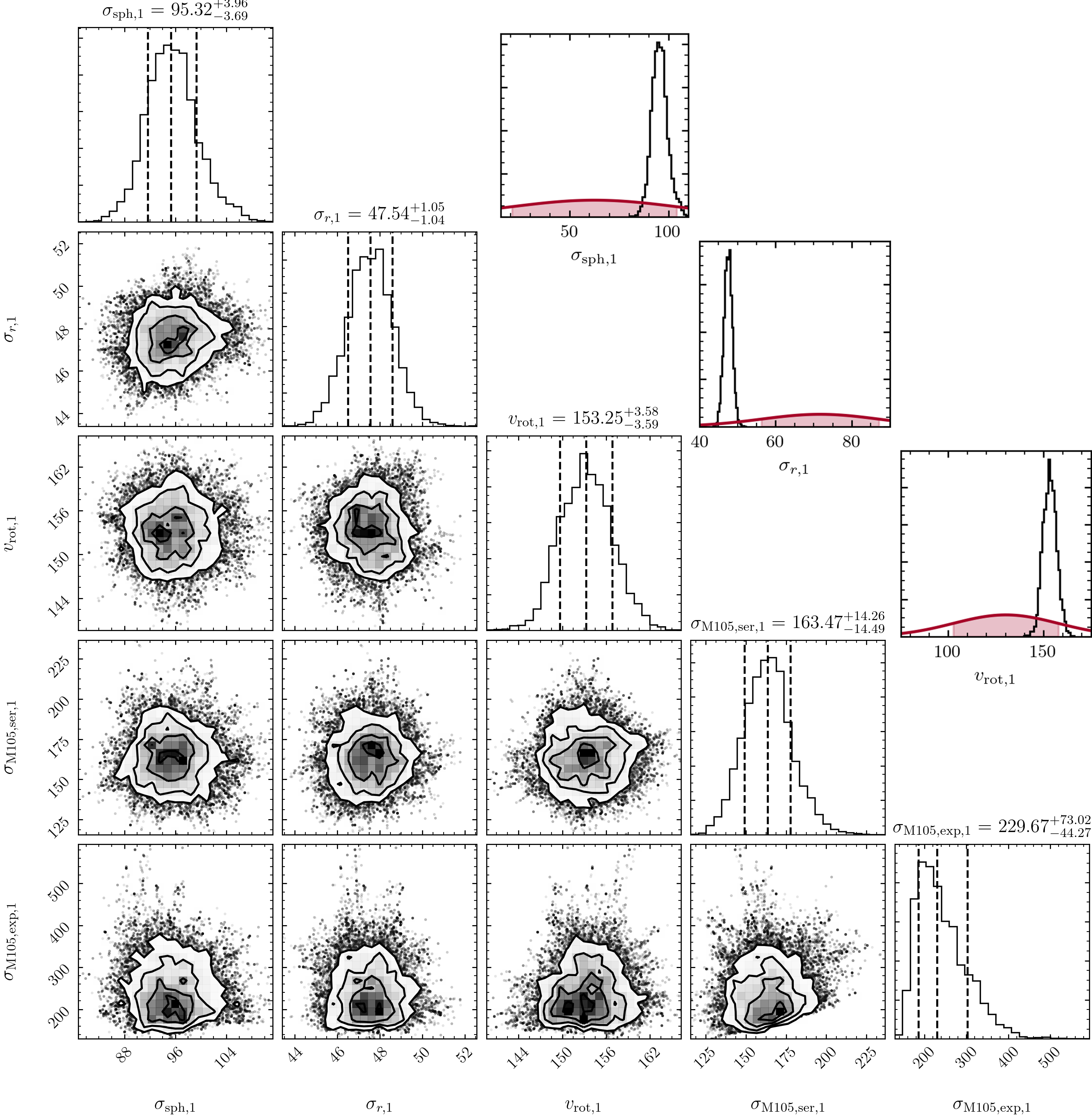}
    \caption{Posterior probability distribution of the parameters $\sigma_\mathrm{sph}$ (the velocity dispersion of the bulge of NGC~3384), $\sigma_r$ and $v_\mathrm{rot}$ (the radial velocity dispersion and circular velocity of the disk of NGC~3384), and $\sigma_\mathrm{M105,Ser}$ and $\sigma_\mathrm{M105,Exp}$ (the velocity dispersions of the S\'{e}rsic and exponential populations in M105) in the innermost of the five elliptical bins. To illustrate the Gaussian priors on the LOSVD parameters of NGC~3384, which cover a larger range in parameter space, we inserted three additional subplots showing the priors with red solid lines and the posteriors with black histograms, utilising the same binning as the corner plots. The red shaded regions below the Gaussians denote $1\sigma$ from the mean. On all plots, the velocities or velocity dispersions are in units of $[\mathrm{km}\,\mathrm{s}^{-1}]$.}
    \label{fig:cornerplot_bin1}
\end{figure*}
As \citet{2013MNRAS.432.1010C} already carried out a bulge-disk decomposition of NGC~3384 based on the single PN.S field available to them at the time, we use their best-fit parameters and corresponding errors as Gaussian priors on $v_\mathrm{rot}, \sigma_r, \sigma_{\phi}, \sigma_\mathrm{N3384,bulge}$. As we evaluate the LOSVD in five elliptical bins instead of three, we interpolated their best-fit profiles as a function of radius and used the interpolated values at the bin centres. An example of the Gaussian priors for the innermost of the five elliptical bins is indicated with red solid lines in Fig.~\ref{fig:cornerplot_bin1}. We place uniform priors on $\sigma_\mathrm{M105,Ser}$ and $\sigma_\mathrm{M105,Exp}$, but require $\sigma_\mathrm{M105,Ser} \leq \sigma_\mathrm{M105,Exp}$ in each of the bins. 

The posterior probability distributions are calculated with \textsc{emcee} \citep{2013PASP..125..306F}, using 52 walkers ($2 N_\mathrm{dim} + 2$) and $2\,000$ steps. Figure~\ref{fig:cornerplot_bin1} shows the resulting distributions visualised as a corner plot for the innermost of the five elliptical bins. This is a convenient representation of a subset of the parameter space. However, we stress that the whole $25$-dimensional space is explored simultaneously for data in the whole radial range. The best-fit parameters and their errors are the 50\% and (16\%,84\%) quantiles. While the best-fit parameters give a good first estimate on the LOS velocity dispersion of the S\'{e}rsic and Elliptical PN populations in the halo of M105, we will calculate more accurate radial velocity dispersion profiles with robust methods and based on the final membership assignment presented in the following.

\subsection{Final membership assignment}
\label{ssec:likelihood}
\begin{figure*}
  \centering
  \includegraphics[width=18cm]{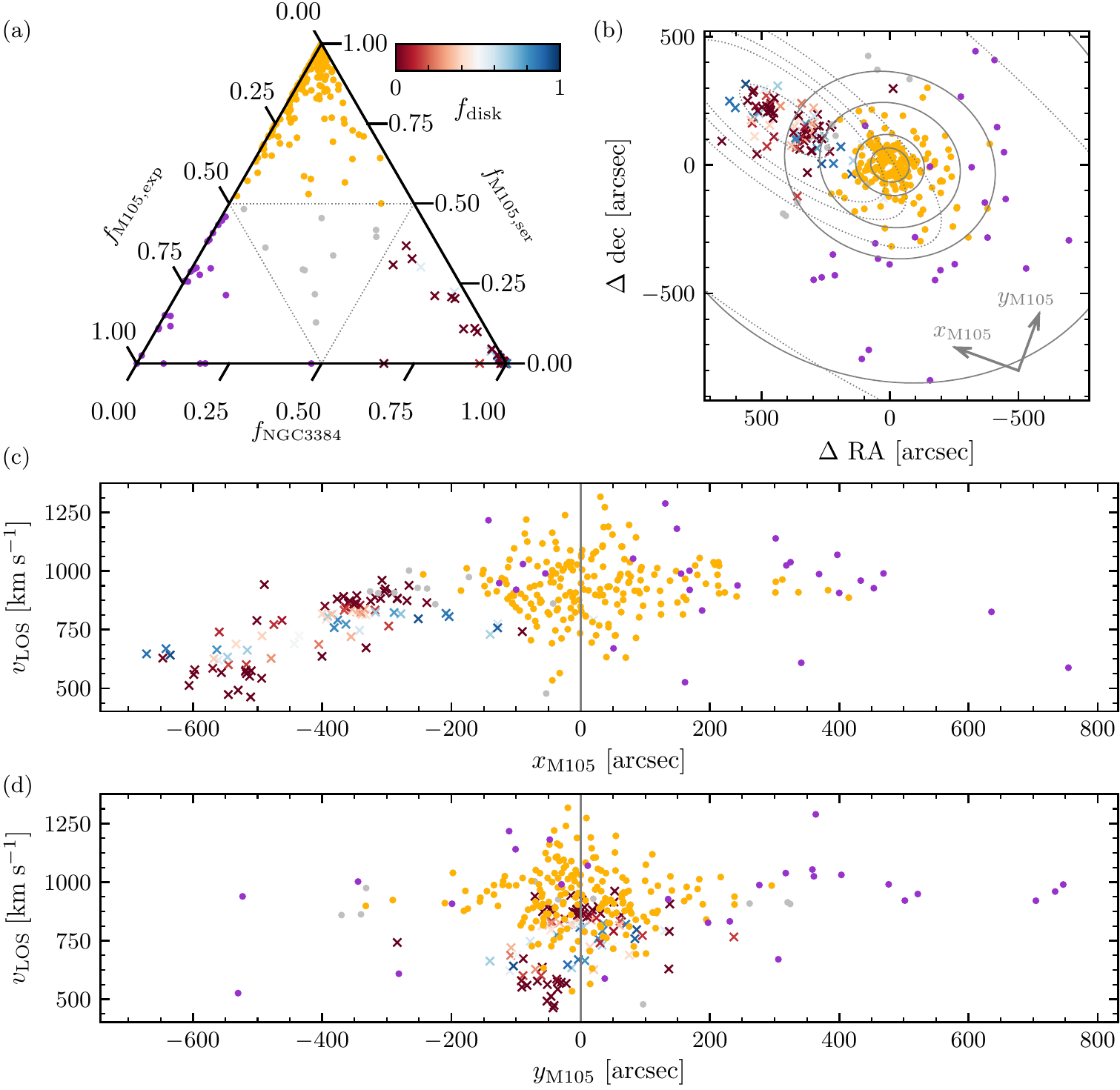}
  \caption{\textit{Panel~(a)} Ternary plot of the probabilistic fractions $f_i$ with $f_\mathrm{M105,Ser} > 0.5$ coloured in orange, $f_\mathrm{M105,Exp} > 0.5$ coloured in purple, and $f_\mathrm{NGC3384} > 0.5$ represented by crosses on a red-to-blue colour scale with red colours for $f_\mathrm{NGC3384,bulge} > 0.5$ and blue colours for $f_\mathrm{NGC3384,disk} > 0.5$. Points coloured in grey do not satisfy any of the four criteria, and their firm association with any of the components cannot be conclusively ascertained. \textit{Panel~(b)} Spatial distribution of PNe with the same colour-coding as in panel (a). \textit{Panel~(c)} Velocity phase-space along the major axis of M105 with the same colour-coding as in panel (a). \textit{Panel~(d)} Velocity phase-space along the minor axis of M105 with the same colour-coding as in panel (a).}
  \label{fig:decomposition}
\end{figure*}
For each PN, we determine the probabilistic fraction $f_i$ to belong to one of the four populations (NGC~3384's disk and bulge, M105's S\'{e}rsic and exponential halos) based on its position and LOS velocity. Figure~\ref{fig:decomposition}(a) illustrates the probability space populated by our observations, with $f_\mathrm{M105,Ser} > 0.5$ coloured in orange (138 PNe), $f_\mathrm{M105,Exp} > 0.5$ coloured in purple (31 PNe), and $f_\mathrm{NGC3384} > 0.5$ being represented by crosses on a red-to-blue colour scale with red colours for $f_\mathrm{NGC3384,bulge} > 0.5$ (37 PNe) and blue colours for $f_\mathrm{NGC3384,disk} > 0.5$ (56 PNe). Points which cannot be uniquely associated with any of the four components outlined previously, are colour-coded in grey (7 PNe).
Figure~\ref{fig:decomposition} (b) shows the distribution of PNe colour-coded by $f_i$ in position-space, and Fig.~\ref{fig:decomposition}(c) and (d) in phase-space along the major and minor axes of M105, respectively. 

Both in position- and phase-space, PNe can be robustly separated into those associated with M105 and those with NGC~3384. This is important for the study of the LOSVD of M105 that will be presented in the following sections. We have maximised the number of PNe that will be considered for follow-up analysis compared to conventional cuts in position space \citep[as for example done by ][]{2007apj...664..257d}, while ensuring that the derived LOSVDs are of sufficient quality. The final column of Tab.~\ref{tab:pns_final} provides the probabilistic fraction to be assigned to M105 $f_\mathrm{M105} = f_\mathrm{M105,Ser} + f_\mathrm{M105,Exp}$ for each PN in the final sample. This can be easily translated to the probabilistic fraction to be assigned to NGC~3384 $f_\mathrm{NGC3384} = 1 - f_\mathrm{M105}$.

\section{Kinematics in the outer halo of M105}
\label{sec:kinematics}
In the following section, we only consider PNe that we assigned to the M105 halo and envelope in the previous section. For the calculation of the smoothed 2D velocity and velocity dispersion fields and the extraction of the rotation profiles, we assume that M105 is point-symmetric in the position-velocity phase-space, as commonly done when constructing velocity fields based on PN kinematics \citep[e.g.][]{1998apj...507..759a, 2009MNRAS.394.1249C, 2018A&A...618A..94P}. This implies that each point $(x,y,v)$ in said phase-space has a "mirror point" $(-x,-y,-v)$ and that the number of data points on which the kinematic quantities are calculated is doubled. The resulting catalogue, i.e. the concatenation of the \emph{original} and \emph{mirrored} catalogues, is called the \emph{folded} catalogue.

\subsection{Smoothed 2D velocity and velocity dispersion fields}
\label{ssec:smooth}

\begin{figure*}
  \includegraphics[width=18cm]{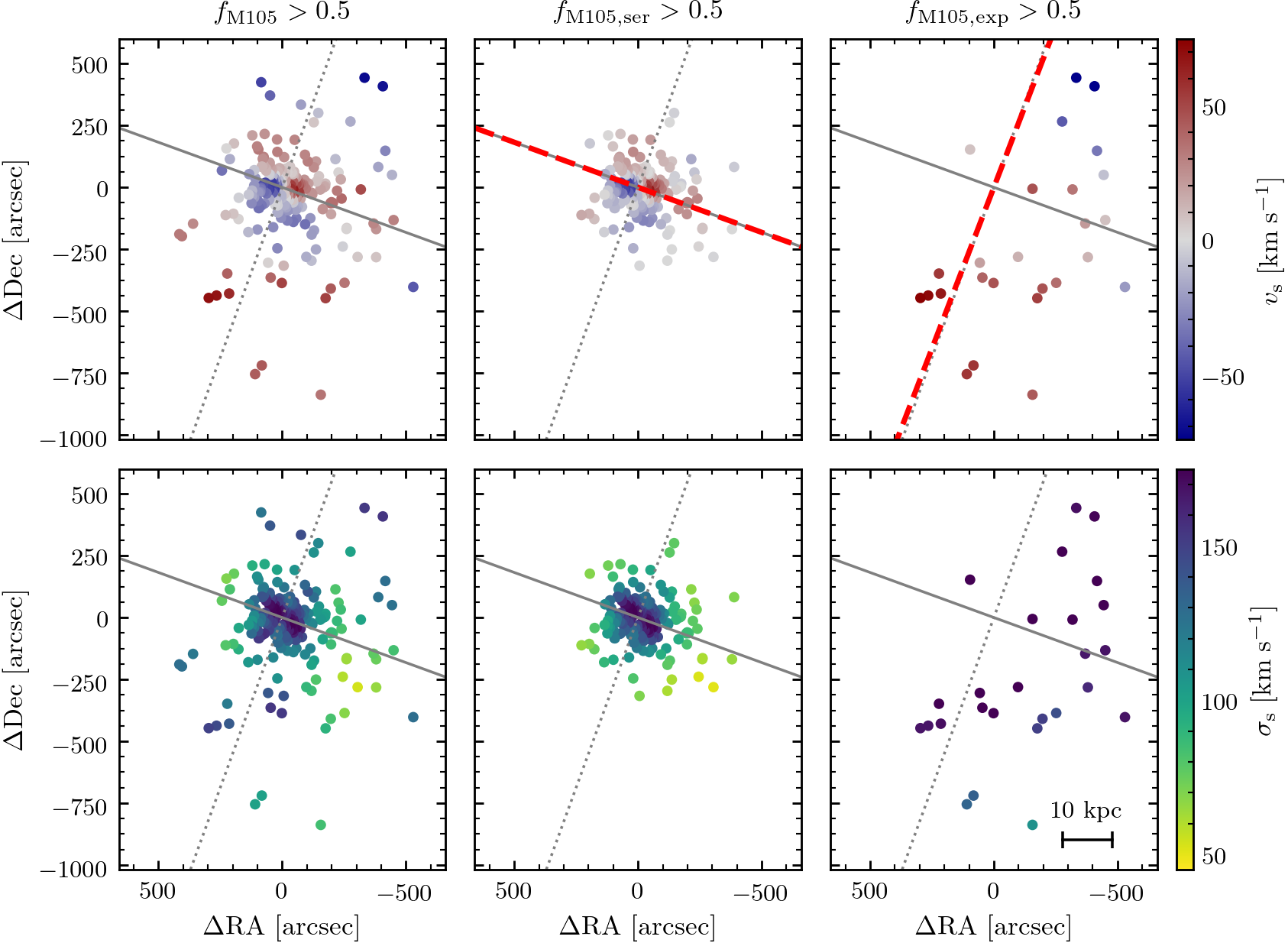}
  \caption{Kernel-smoothed velocity (top row) and velocity dispersion fields (bottom row) for PNe associated with M105 at large (left column), the S\'{e}rsic component (middle column), and the exponential component (left column). Each point denotes the position of an observed PN, colour-coded by the value of the smoothed field. The solid (dotted) grey lines denote the photometric major (minor) axis and the dashed red line the kinematic major axis. North is up, and East is to the left.}
  \label{fig:velfields}
\end{figure*}

We calculate kernel-smoothed velocity and velocity dispersion fields for all PNe associated with M105 and for the S\'{e}rsic and exponential components separately from the folded catalogues. We use a distance-dependent Gaussian kernel as defined in \citet{2009MNRAS.394.1249C}
\begin{equation}
  w_i = \exp\left(\frac{-D_i^2}{2k(x_i,y_i)^2}\right),
\end{equation}
where the kernel width $k(x,y)$ is proportional to the distance $R_{i,M}$ to the $M$th closest tracers located at $(x_M, y_M)$:
\begin{equation}
  k(x_i,y_i) = A R_{i,M}(x_M, y_M) + B.
\end{equation}
We use the optimised kernel parameters $A = 0.34$ and $B = 16.2$ for the $M=20$ closest tracers, which were determined by \citet{2018A&A...618A..94P} based on Monte Carlo simulations of discrete velocity fields, allowing for the best compromise between noise smoothing and spatial resolution. 

Figure~\ref{fig:velfields} shows the resulting smoothed velocity $v_\mathrm{s}$ (top row) and velocity dispersion $\sigma_\mathrm{s}$ fields (bottom row) for all PNe associated with M105 in the left column, and the S\'{e}rsic and exponential components shown in the middle and right columns, respectively. 
The smoothed velocity and velocity dispersion fields of the S\'{e}rsic and exponential components show distinct features. As expected from the method with which PNe were associated to either component, the LOS velocity dispersion of PNe in the exponential component is much larger than that of those in the S\'{e}rsic one, even at comparable radial ranges. The rotation signatures in the velocities are also different, hinting towards different mechanisms that contributed to the formation of the S\'{e}rsic and exponential components in the halo of M105. 

\subsection{Line-of-sight velocity dispersion profiles}
\label{sec:losvd}
\begin{figure*}
  \includegraphics[width=18cm]{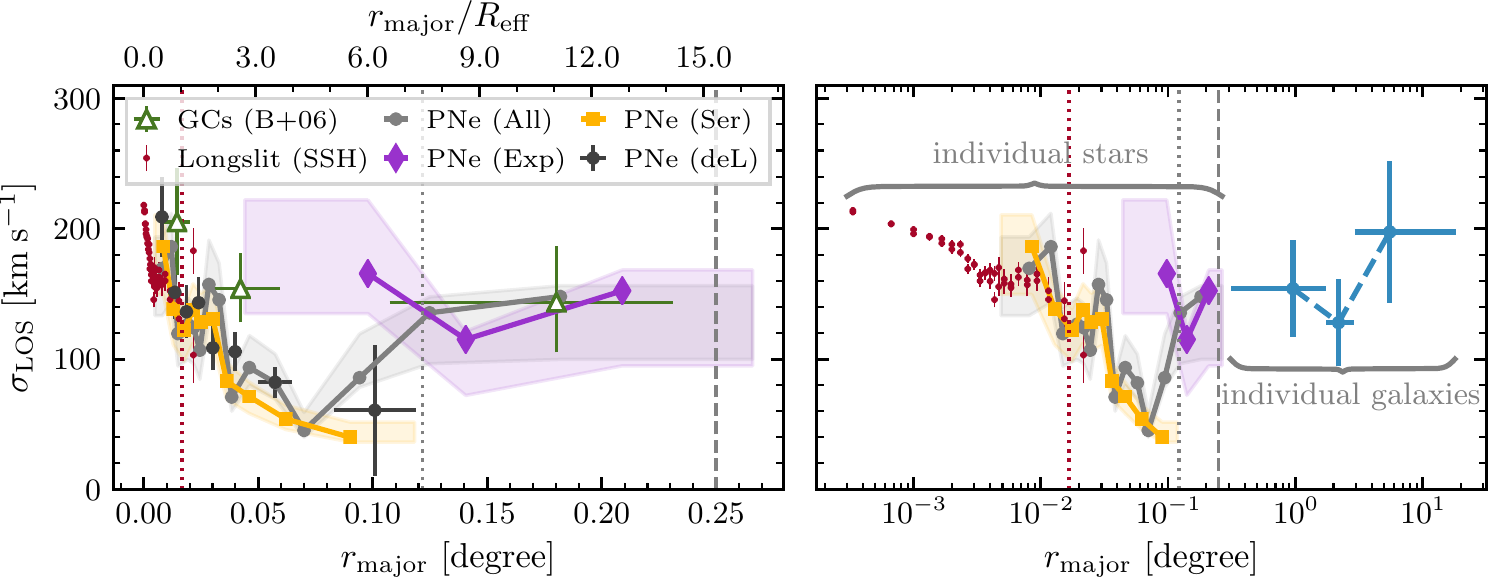}
  \caption{\textit{Left:} LOS velocity dispersion profiles of M105 calculated with robust techniques for the total sample of PNe associated with M105 (solid lines with grey dots), its S\'{e}rsic component (solid lines with orange squares), and the exponential envelope (solid lines with purple diamonds). The transparent bands denote the errors on the derived LOS velocity dispersions estimated with robust methods. For comparison, we also show the LOS velocity dispersion used by \citet[][black error bars]{2009mnras.395...76d} and the one derived from long-slit spectroscopy \citep[small red error bars][]{1999AJ....117..839S}, as well as the globular cluster (GC) velocity dispersion from \citet[][green triangles with errorbars]{2006A&A...448..155B}. 
  \textit{Right:} As the left panel, but with a logarithmic scale for the major-axis radius. The blue error bars denote the velocity dispersion calculated from the LOS velocities of satellites in the Leo~I group \citep[compiled by][]{2018A&A...615A.105M}. In both panels, shaded regions indicate the robust errors, and the vertical lines are the same as in Fig.~\ref{fig:rotation_rmajor}.}
  \label{fig:M105sigmas}
\end{figure*}
To determine the LOS velocity dispersion profiles, we again only consider PNe with $f_\mathrm{M105} = f_\mathrm{M105,Ser} + f_\mathrm{M105,Exp} > 0.5$, i.e. PNe firmly associated with M105 and the exponential envelope both in position and velocity space. 
We already obtained velocity dispersions in the five elliptical bins from the likelihood analysis presented in Sect.~\ref{ssec:likelihood} for the S\'{e}rsic and exponential components separately. However, to explore a finer resolution along the radial direction and be more robust to potential outliers, we re-bin the data and use robust estimation methods \citep{2010a&a...518a..44m} with a clipping range of $2\sigma$. 

Figure~\ref{fig:M105sigmas} shows the resulting LOS velocity dispersion for all PNe associated with M105 with grey dots and the corresponding uncertainty with a grey error band. The overall velocity dispersion agrees well with that presented in \citet[][black error bars]{2009mnras.395...76d} where the data overlaps, as well as with long-slit data in the innermost region of the galaxy from \citet{1999AJ....117..839S}. 

The velocity dispersion increases strongly beyond $400\arcsec$ (corresponding to $\approx 7.5 R_\mathrm{eff}$), which also corresponds to the radius at which the PN number density starts to flatten due to a higher $\alpha$-parameter value at large radii, as indicated by the dotted vertical line. 
The rise in the LOS velocity at large radii is thus driven by PNe associated with the exponential envelope (indicated by purple triangles on Fig.~\ref{fig:M105sigmas}). These PNe have a larger LOS velocity dispersion compared to those associated with the S\'{e}rsic halo (orange squares). The strong decline of the LOS velocity dispersion in the inner halo is instead driven by PNe associated with the latter component.

\subsection{Rotation}
\begin{figure}
  \includegraphics[width=8.8cm]{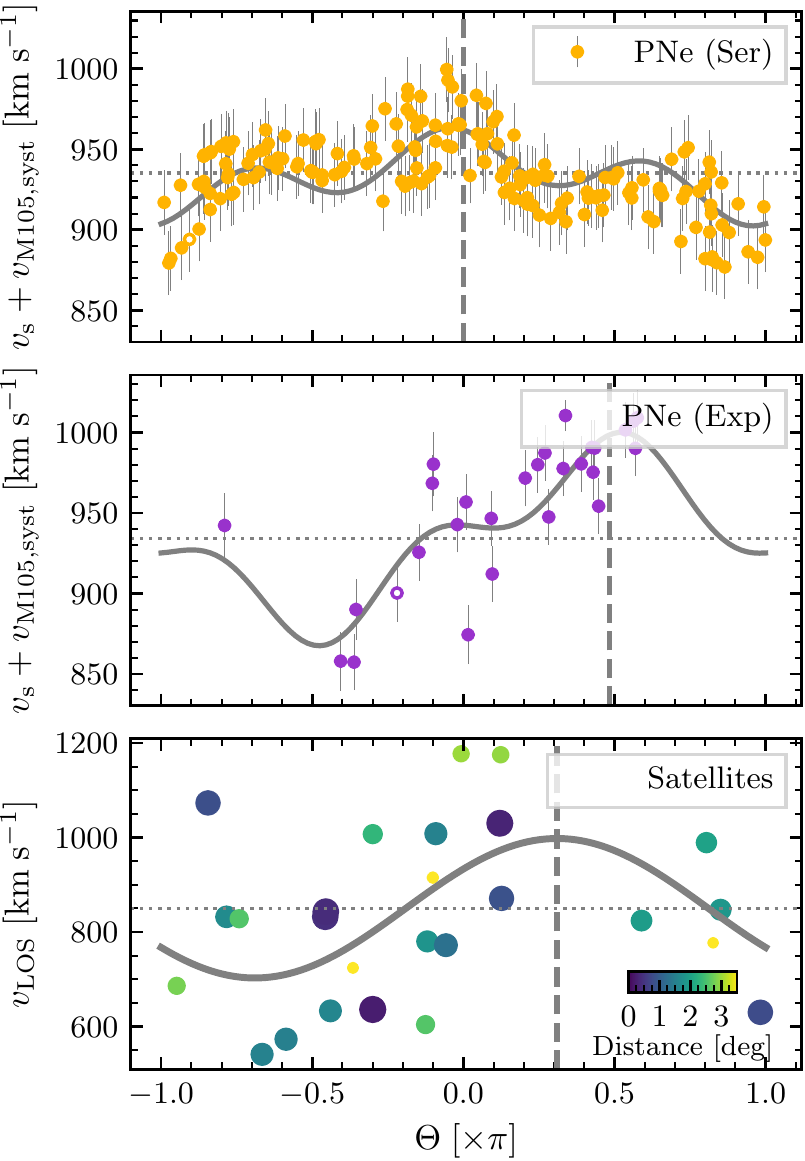}
  \caption{Radially integrated rotation profiles of PNe associated with the S\'{e}rsic (\textit{top panel}) and exponential (\textit{middle panel}) populations from the kernel-smoothed velocity maps (dots with error bars). The \textit{bottom panel} shows unsmoothed LOS velocities of satellite galaxies in the Leo~I group colour-coded by their on-sky distance to the centre of M105 \citep[compiled by][]{2018A&A...615A.105M}. In each panel, the best-fit rotation profiles are denoted with solid grey curves, and the corresponding best-fit kinematic position angles (with respect to the photometric one) and systemic velocities are denoted with dashed and dotted grey lines. Open symbols denote PNe excluded from the fit as they lie outside of the $2\sigma$ limits determined in Sect.~\ref{sec:losvd}.}
  \label{fig:rotation}
\end{figure}

We first assess the different rotation signatures, by separately fitting rotation curves to the smoothed velocities $v_s$ for the S\'{e}rsic and exponential populations separately as a function of position angle $\Theta$ as shown on Fig.~\ref{fig:rotation}, excluding the 2$\sigma$ outliers identified in the previous Sect.~\ref{sec:losvd} from further analysis. Following \citet{2018A&A...618A..94P}, we fit the mean velocity fields with the following function:
\begin{align}
\begin{split}
      v_p(R,\Theta) &= v_\mathrm{sys}(R) + v_\mathrm{rot}(R) \cos(\Theta - \Theta_0(R)) \\
      &+  s_3(R) \sin(3\Theta - 3\Theta_0) + a_3(R) \cos(3\Theta - 3\Theta_0),
      \label{eqn:rotation}
\end{split}
\end{align}
where $v_\mathrm{sys}$ is the systemic velocity, $v_\mathrm{rot}$ the amplitude of the rotation, and $\Theta_0$ the kinematic position angle with respect to the photometric position angle. These parameters describe the rotation around the kinematic major axis, while $s_3$ and $a_3$ are the amplitudes of the third-order terms.\footnote{As we assume that the phase-space is point-symmetric, this prohibits even contributions to the higher-order terms.}

Figure~\ref{fig:rotation} shows the smoothed velocity as function of the position angle $\Theta$ for the observed PNe in each of the two components. We fit rotation profiles to these data, only considering the azimuthal dependence of eq.~\eqref{eqn:rotation}. We evaluate whether the goodness of the fit is improved by including the higher-order terms with amplitudes $a_3$ and $s_3$ by virtue of the Bayesian information criterion (BIC). 
Table~\ref{tab:rotation} summarises the best-fit parameters fit to the velocity fields, and the resulting best-fit rotation curves are denoted with solid grey lines. Models including higher moments are preferred for both components. In the case of the S\'{e}rsic component, the fit with higher moments has both lower BIC and reduced $\chi^2$ values. For the exponential envelope, the BIC values are indistinguishable, but the fit with higher moments has a lower reduced $\chi^2$ value.

\begin{table*}[]
    \centering
    \caption{Best-fit parameters for the rotation models fit to the S\'{e}rsic and exponential velocity fields.}
    \begin{tabular}{lllllrrrr}
    \hline 
    \hline
    Component & Higher modes & $\Theta_0$\tablefootmark{(a)} &$v_\mathrm{sys}$\tablefootmark{(b)} & $v_\mathrm{rot}$ & \multicolumn{1}{l}{$s_3$} & \multicolumn{1}{l}{$a_3$} & \multicolumn{1}{l}{BIC} & \multicolumn{1}{l}{$\chi^2_\mathrm{red}$} \T \\
     & & $[\times\pi]$ & $[\mathrm{km\,s^{-1}}]$ & $[\mathrm{km\,s^{-1}}]$ & \multicolumn{1}{l}{$[\mathrm{km\,s^{-1}}]$} & \multicolumn{1}{l}{$[\mathrm{km\,s^{-1}}]$} & & \B \\
     \hline 
     S\'{e}rsic & Y & $0.00 \pm 0.13$ & $935.1 \pm 1.4$ & $15.9 \pm 1.9$ & $-6.46\pm5.66$ & $13.3 \pm 3.1$ & -10.6 & 153.6\T\\
     S\'{e}rsic & N & $0.00 \pm 0.02$ & $933.7 \pm 1.5$ & $15.5 \pm 2.0$ & & & 28.4 & 200.2\\
     Exponential & Y & $0.48 \pm 0.09$ & $934.2 \pm 9.9$ & $49.3 \pm 8.1$ & $8.79 \pm 14.30$ & $16.3 \pm 8.4$ & 30.0 & 44.1 \\
     Exponential & N & $0.56 \pm 0.09$ & $925.9 \pm 9.0$ & $49.8 \pm 8.0$ & & & 29.8 & 56.1\B\\
     \hline 
     \hline
    \end{tabular}
    \tablefoot{
  \tablefoottext{a}{The kinematic position angle was determined with respect to the photometric position angle.}
  \tablefoottext{b}{The systemic velocity determined by \citet{2018A&A...618A..94P} based on the M105-C field is $v_\mathrm{sys,M105} = 934\pm14\;\mathrm{km\,s^{-1}}$.}
  }
    \label{tab:rotation}
\end{table*}

The S\'{e}rsic component has a rotation amplitude of $v_\mathrm{rot,Ser} = 16.0 \pm 1.8\;\mathrm{km}\,\mathrm{s}^{-1}$ and a kinematic major axis that is consistent with the photometric major axis within the errors, as well as a systemic velocity of $v_\mathrm{sys,Ser} = 935.1\pm1.4\;\mathrm{km}\,\mathrm{s}^{-1}$ that agrees with that determined by \citet{2018A&A...618A..94P} based on the central field of M105.
In contrast to this, the rotation amplitude of the PNe in the exponential envelope is more than three times higher, being $v_\mathrm{rot,Exp} = 49.2 \pm 6.7\;\mathrm{km}\,\mathrm{s}^{-1}$, and the kinematic major axis is aligned with the photometric minor axis of M105, while their systemic velocity of $v_\mathrm{sys,Exp} = 934.2\pm9.9\;\mathrm{km}\,\mathrm{s}^{-1}$ agrees with that of the S\'{e}rsic component within the errors. To visualise the on-sky orientation of the photometric and kinematic major axes, they are overplotted on top of the smoothed velocity fields in Fig.~\ref{fig:velfields} with solid grey and dashed red lines, respectively. 

\begin{figure}
    \centering
    \includegraphics[width=8.8cm]{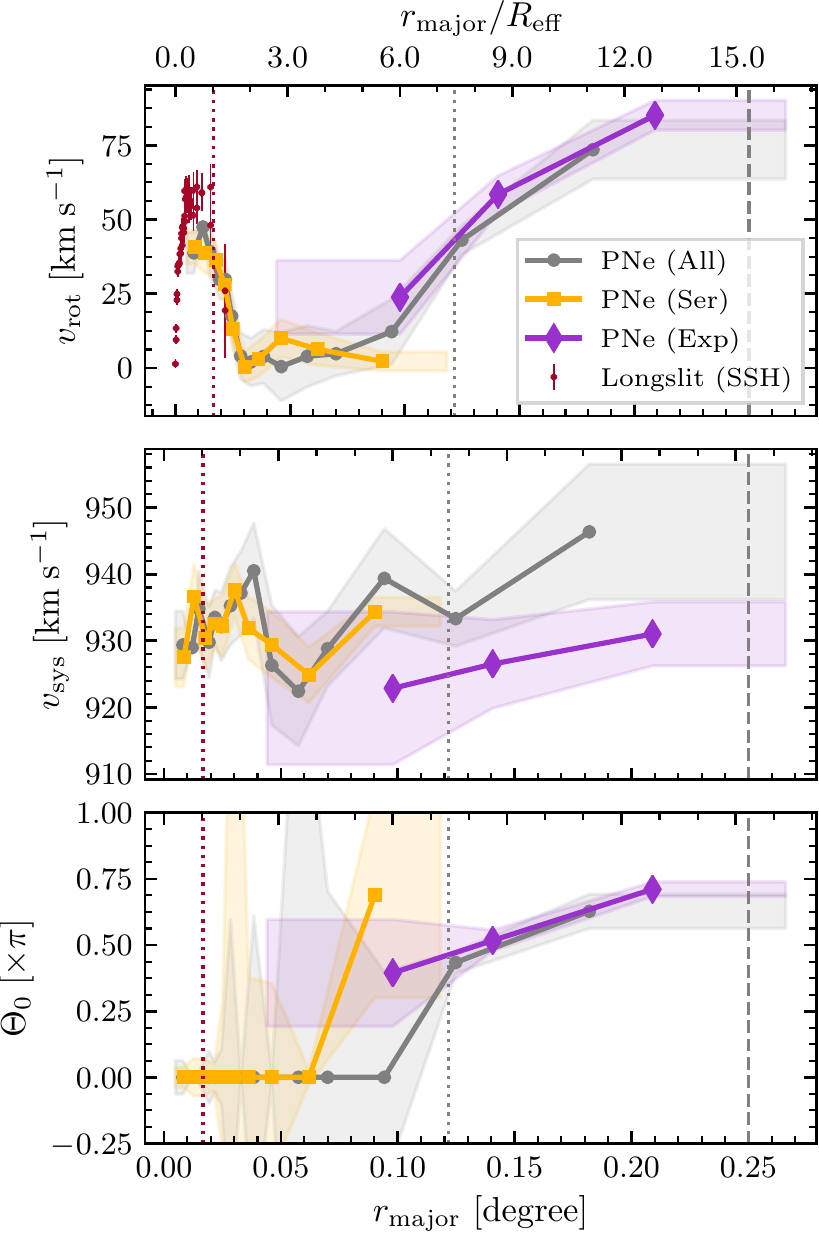}
    \caption{\textit{Top:} Rotation amplitude along the major axis of M105 for all PNe associated with the galaxy (grey circles), and the S\'{e}rsic (orange squares) and exponential (purple diamonds) components. The red points with associated error bars denote the rotation profile derived from long-slit spectroscopy \citep{1999AJ....117..839S}. \textit{Middle:} Systemic velocity as function of majr-axis radius with the same colour-coding as in the top panel. \textit{Bottom:} Best-fit kinematic position angle with respect to the photometric one as function of major-axis radius with the same colour coding as in the top panel. In both panels, shaded regions indicate the standard errors from the maximum-likelihood fit, the dotted grey lines denote the radius at which the $\alpha$-parameter changes, and the grey dashed vertical lines where the exponential SB starts to dominate the total light distribution. The dotted red vertical line denotes the kinematic transition radius determined by \citet{2018A&A...618A..94P}.}
    \label{fig:rotation_rmajor}
\end{figure}

To evaluate the radial dependence of the rotation profile, we again fitted eq.~\ref{eqn:rotation} to the data, but in elliptical bins, for all PNe associated with M105 as well as to the S\'{e}rsic and exponential populations separately. Due to the smaller number of tracers per bin, we did not fit the third-order modes, as this led to noisier fits. The resulting best-fit profiles of the rotation amplitude $v_\mathrm{rot}$, systemic velocity $v_\mathrm{sys}$ and kinematic position angle $\theta_0$ are shown in Fig.~\ref{fig:rotation_rmajor} from top to bottom. As already observed by \citet{2018A&A...618A..94P}, the rotation amplitude decreases in the inner halo with hints for growing rotation in the outskirts. This decrease was also inferred from long-slit spectroscopy \citep{1999AJ....117..839S}. Our new extended data reveal that there is a transition region where the rotation amplitude remains small, followed by a strong increase of the rotation amplitude in the exponential envelope which is driven by the PNe associated with this component (purple diamonds). 

This transition goes hand in hand with a twisting of the kinematic position angle that is illustrated in the bottom panel of Fig.~\ref{fig:rotation_rmajor}. In the inner halo, where the S\'{e}rsic population dominates, the kinematic position angle is aligned with the photometric one. In contrast, in the outer halo, the position angle twists by more than 90\degr and is more or less aligned with the photometric minor axis of the inner high SB regions.

\subsection{Angular-momentum content}
\begin{figure*}
    \centering
    \includegraphics[width=18cm]{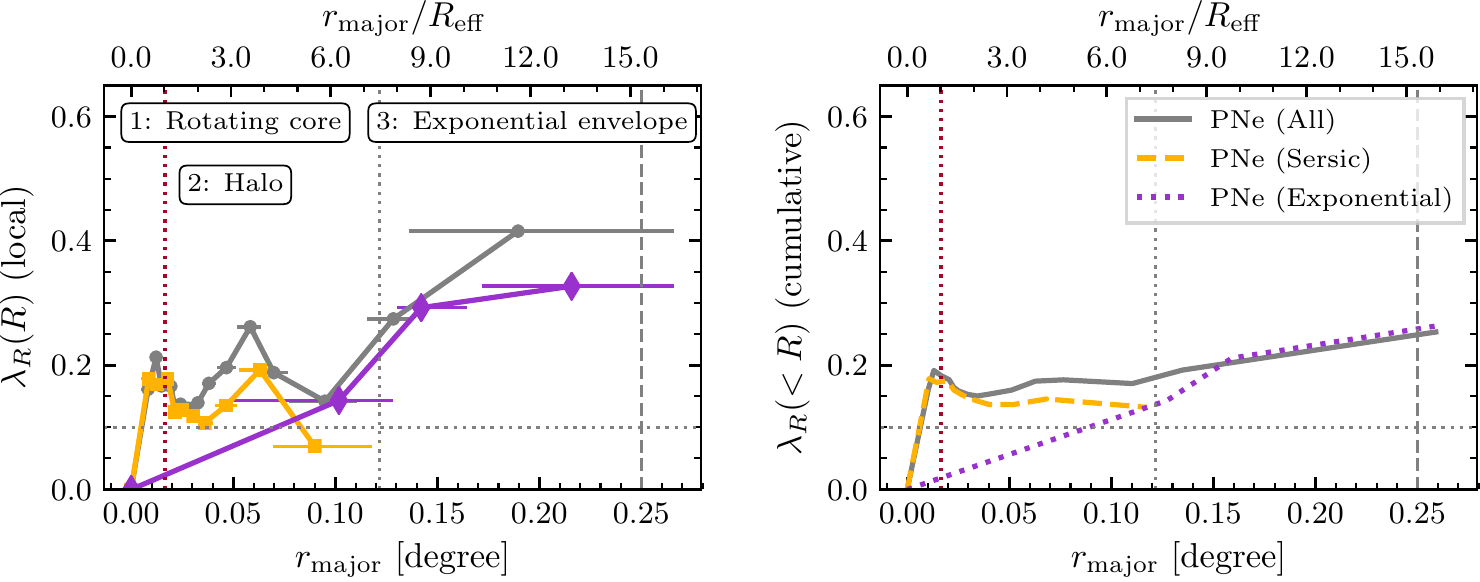}
    \caption{\textit{Left:} Local $\lambda_R$ profiles calculated for the total sample of PNe associated with M105 (grey), its S\'{e}rsic component (orange), and the exponential envelope (purple). Horizontal error bars denote the bins in which $\lambda_R$ was evaluated. \textit{Right:} Cumulative $\lambda_R$ profiles, with the same colour-coding as in the left panel. In both panels, the dotted grey line at $\lambda_R$ = 0.1 separates fast from slow rotator regions assuming constant ellipticity. \citep{2007MNRAS.379..401E}. The vertical dotted lines are the same as in Fig.~\ref{fig:rotation_rmajor}.}
    \label{fig:lambda}
\end{figure*}
\citet{2007MNRAS.379..401E} introduced a kinematic classification scheme for galaxies based on the so-called $\lambda_R$ parameter, which quantifies rotational support and can be used as a proxy to quantify the observed projected stellar angular momentum. To infer the angular momentum content locally, i.e. in elliptical bins with mean radius $R_\mathrm{mean}$ and bin edges $R_\mathrm{min}$ and $R_\mathrm{max}$, $\lambda_R$ can be calculated from the smoothed velocity and velocity dispersion fields $V$ and $\sigma$:
\begin{equation}
    \lambda_R(R_\mathrm{mean}) = \frac{\sum_{R_i=R_\mathrm{min}}^{R_\mathrm{max}} R_i \left| V_i \right| }{\sum_{R_i=R_\mathrm{min}}^{R_\mathrm{max}} R_i \sqrt{V_i^2 + \sigma_i^2}}.
\end{equation}
The cumulative $\lambda_R$ profiles instead are calculated from the centre of the galaxy to the outer bin edge $R_\mathrm{max}$ and corrected for geometric incompleteness $c_\mathrm{geo}$:
\begin{equation}
    \lambda_R(< R_\mathrm{max}) = \frac{\sum_{R_i=0}^{R_\mathrm{max}} R_i \left| V_i \right|/c_{\mathrm{geo}} }{\sum_{R_i=0}^{R_\mathrm{max}} R_i/c_{\mathrm{geo}} \sqrt{V_i^2 + \sigma_i^2}}.
\end{equation}
The cumulative $\lambda_R$ is weighted by the flux associated with each radial bin. For PNe, this is implicitly incorporated as the PN number density traces the light distribution \citep{2009MNRAS.394.1249C}.
When ordered motion, i.e. rotation, dominates, $\lambda_R$ approaches unity. Within one effective radius, $\lambda_R$ can be used as a proxy to divide galaxies into the categories of fast ($\lambda_R > 0.1$) and slow rotators ($\lambda_R < 0.1$). M105 has been classified as a fast rotator \citep{2007MNRAS.379..401E, 2009MNRAS.394.1249C, 2018A&A...618A..94P}.

The left panel of Fig.~\ref{fig:lambda} shows the local $\lambda_R$ profiles for all PNe associated with M105 (grey), the S\'{e}rsic halo (orange) and the exponential envelope (purple), evaluated in the same elliptical annuli as used in Figs.~\ref{fig:rotation_rmajor} and \ref{fig:M105sigmas}. In the inner halo, the kinematic transition radius identified by \citet[][dotted vertical red line on Fig.~\ref{fig:lambda}]{2018A&A...618A..94P} is marked by a decrease in the local $\lambda_R$ profile, which pertains until $\sim 2R_\mathrm{eff}$. At $\sim 4R_\mathrm{eff}$ the $\lambda_R$ profile peaks again. At large radii, i.e. in the exponential envelope, $\lambda_R$ increases, reaching values of $\lambda_R=0.4$ in the outermost bin. The right panel of Fig.~\ref{fig:lambda} shows the cumulative $\lambda_R$ profiles with the same colour coding. Based on the two panels, we identify three distinct kinematic components, which will be discussed further in the following section:
\begin{enumerate}
    \item the \emph{rotating core} within $1~R_\mathrm{eff}$,
    \item the \emph{halo}, from $1~R_\mathrm{eff}$ to $7.5~R_\mathrm{eff}$, and 
    \item the \emph{exponential envelope}, from $7.5~R_\mathrm{eff}$ to the last data point at $16~R_\mathrm{eff}$.
\end{enumerate}

\section{Discussion}
\label{sec:discussion}
In this work, we have efficiently associated PNe in the Leo~I group to different subpopulations in the velocity phase-space centred on M105. Vital for measuring the LOS kinematics at large radii from the centre of M105 was the division of the sample into PNe associated with M105, the exponential envelope, and with the companion galaxy NGC~3384. We refer the reader to \citet{2013a&a...549a.115c, 2013MNRAS.432.1010C} for a discussion of the kinematics of the S0 galaxy NGC~3384 and discuss the kinematics of M105 and the surrounding IGL in what follows.

\subsection{Metal-rich and intermediate-metallicity populations in the inner halo of M105}
\citet{2018A&A...618A..94P} identified a kinematic transition in the inner halo at $1~R_\mathrm{eff}$, which is marked by a decrease of the rotation amplitude  (Fig.~\ref{fig:rotation_rmajor}), the LOS velocity dispersion (Fig.~\ref{fig:M105sigmas}), and the local $\lambda_R$ parameter (Fig.~\ref{fig:lambda}) with radius. 
The first kinematic component of M105 that we identified is thus the \emph{rotating core} in the centre of the galaxy. Our measurements of $V$ and $\sigma$ agree with those of \citet{2009mnras.398..561w} based on SAURON integral-field spectroscopy in the regions of overlap and the long-slit data from \citet{1999AJ....117..839S}. \citet{2009mnras.398..561w} inferred an age of $\approx 12$ Gyr and approximately solar metallicity for the rotating core. 

Beyond the transition at $1~R_\mathrm{eff}$, the metallicity decreases, reaching 20\% of the solar value, i.e. $\approx -0.7$ at $3-4~R_\mathrm{eff}$ \citep{2009mnras.398..561w}, which is similar to the peak of the metallicity distribution function in the inner \textit{HST} field located $4 \lesssim R_\mathrm{eff} \lesssim 6$ to the northeast of the centre of M105 \citep[][see also the hatched regions in the right panel of Fig.~\ref{fig:survey}]{2016ApJ...822...70L}. The PN population in the inner halo is characterised by a low $\alpha$-parameter $\alpha_\mathrm{2.5,MR+IM} = (1.00 \pm 0.11)\times 10^{-8}\;\mathrm{PN}\,L_\mathrm{bol}^{-1}$ and a shallow, but still steeper than the ``standard'' \citet{1989apj...339...53c} PNLF slope \citepalias{2020A&A...642A..46H}. 

In the $\lambda_R$ profile shown in Fig.~\ref{fig:lambda}, a second peak is visible at around $4~R_\mathrm{eff}$. This peak is due to an increase in rotation at these radii, while the LOS velocity dispersion profile monotonously declines in the inner halo. This peak is co-spatial with a change in position angle of the isophotes of M105 and an increased ellipticity (VEGAS collaboration, priv. communication) and may be related to a secondary rotating component, reviving the original suggestion of \citet{1991ApJ...371..535C} that M105 may be a S0 galaxy observed face-on.

Based on results from the IllustrisTNG cosmological hydrodynamical simulations \citet{2021A&A...647A..95P} argued that the observed kinematic transition radii do not trace the transition between in-situ and ex-situ dominated regions. The innermost transition radius at $1~R_\mathrm{eff}$ therefore does not necessarily represent a transition to a component dominated by accreted stars. Instead, the characteristic peaked and outwardly decreasing rotation profile of stars in the inner halo (see top panel of Fig.~\ref{fig:rotation_rmajor}) is similar to that of the in-situ stars in low-mass ETGs in cosmological hydrodynamical simulations \citep{2021A&A...647A..95P}. 

\subsection{The extended, metal-poor envelope in the outer halo of M105 and the IGL of the Leo~I group}
\label{ssec:environment}
The more extended PN.S data in the halo of M105 has allowed us to reveal a second kinematic transition at $\approx 7.5 R_\mathrm{eff}$, where both the LOS velocity dispersion as well as the rotation amplitude increase significantly. This increase is driven by the exponential envelope; a population of PNe associated with a metal-poor population ([M/H] $\leq -1$) of RGB stars following an exponential SB profile \citepalias{2020A&A...642A..46H}. Since M105 is a member of the Leo~I group, we now place these findings into the  context of the kinematics of the group at large.

The increase of the velocity dispersion at large radii may indicate that PNe in the exponential envelope of M105 are bound to the gravitational potential of the Leo~I group and thus part of its IGL. We therefore compare the kinematics in the outer halo of M105 with that of satellite galaxies in the Leo~I group. We use the compilation of \citet[][Tab.~A.4 therein]{2018A&A...615A.105M}, and selected dwarf galaxies in the M96 subgroup with LOS velocity measurements \citep[from][]{1998natur.391..461f, 1992MNRAS.258..334S, 2003A&A...401..483H, 2004AJ....127.2031K, 2004ARep...48..267K, 2011AJ....142..170H, 2013AJ....145..101K}. Unfortunately, the majority of dwarf galaxies in this sample did not have accurate (if any) distance measurements.

With this caveat in mind, we fit the overall rotation profile and determined the LOS velocity dispersion of these 27 galaxies. The bottom panel of Fig.~\ref{fig:rotation} shows their unsmoothed LOS velocities, colour-coded by the on-sky distance to the centre of M105. We fit eq.~\eqref{eqn:rotation} to the data and determine a best-fit systemic velocity of $v_\mathrm{sys,Leo~I} = 850 \pm 35\;\mathrm{km}\,\mathrm{s}^{-1}$ (dotted horizontal line on Fig.~\ref{fig:rotation}), which is lower than that of the PNe in the S\'{e}rsic halo of M105 $v_\mathrm{sys,Ser} = 935.1\pm 1.4\;\mathrm{km}\,\mathrm{s}^{-1}$ and in the exponential envelope $v_\mathrm{sys,Exp} = 934.2\pm 9.9\;\mathrm{km}\,\mathrm{s}^{-1}$. 
The rotation amplitude of the dwarf galaxies $v_\mathrm{rot,Leo~I} = 148\pm 53\;\mathrm{km}\,\mathrm{s}^{-1}$ is significantly larger than that in the outer halo of M105 ($83 \pm 5\;\mathrm{km}\,\mathrm{s}^{-1}$ at the last data point). 
The kinematic position angle of the dwarf galaxies is $PA = 125 \pm 17\degr$ (dashed vertical line), which is nearly aligned with the photometric minor axis of the inner high SB region of M105 and thus also with the kinematic position angle of the exponential envelope, within the errors. 
The best fit is indicated by the solid grey line on the bottom panel of Fig.~\ref{fig:rotation}.

The blue error bars on the right panel of Fig.~\ref{fig:M105sigmas} denote the LOS velocity dispersion of the dwarf galaxies that was determined in three elliptical bins with the same position angle and ellipticity as used when binning the PN data. The LOS velocity dispersion of PNe tracing the exponential envelope (purple diamonds) reaches that of the Leo~I group as traced by the dwarf galaxies. This indicates that both the PNe in the exponential envelope and the surrounding dwarf galaxies trace the group potential. This is corroborated by the similar rotation properties (cf. Fig.~\ref{fig:rotation}).

The increase of LOS velocity dispersion profile at large radii inferred from PNe is corroborated by velocity measurements of globular clusters (GCs). \citet{2006A&A...448..155B} obtained radial velocities of 42 GCs in the Leo~I group, of which they associated 30 with M105, and combined those with previous velocity measurements of eight GCs centred on M105 \citep{2004A&A...415..123P}. The LOS velocity dispersion measurement of \citet{2006A&A...448..155B} is indicated by green triangles with errorbars in the left panel of Fig.~\ref{fig:M105sigmas}. At large radii, the measurements from GCs and PNe are in excellent agreement, while the velocity dispersion measured from GCs in the inner halo is larger than that from PNe. This is expected, since the the GCs have a shallower number density profile than the PNe at these radii \citep{2004A&A...415..123P, 2006A&A...448..155B}. Dividing the sample by colour into blue (and metal-poor) and red (and metal-rich) GCs, \citet{2004A&A...415..123P} find the blue GCs to have a shallower number density profile than the red ones, and in agreement with the measurements from resolved stellar populations \citep{2016ApJ...822...70L} the fraction of blue and metal-poor GCs increases with radius \citep{2006A&A...448..155B}. Because of the small number of GC radial velocities around M105, it is not possible to robustly establish whether the projected rotation of the GCs cospatial with the exponential outer envelope is consistent with that measured using PNe. 

Similar trends of increasing LOS velocity dispersion profiles have been observed for discrete tracers such as PNe and GCs of the IGL or intra-cluster light (ICL) in the Virgo \citep{2018A&A...616A.123H, 2018A&A...620A.111L,  2018ApJ...864...36L} and Fornax Clusters \citep{2018mnras.477.1880s, 2018MNRAS.481.1744P}, as well as based on integrated light in more distant clusters \citep{1979apj...231..659d, 2002apj...576..720k,2015ApJ...807...56B}. While the ICL fraction in these environments is much higher and measured at SB levels larger than that in the Leo~I group, the kinematic signature of halo-to-ICL transition is the same in the low-mass Leo~I group.

\citetalias{2020A&A...642A..46H} argued that the high $\alpha$-parameter value and steeper PNLF slope of the exponential envelope traced by the metal-poor stellar population is indicative for a distinct origin compared to the metal-richer main halo. Furthermore, the  metallicity distribution function of RGB stars in the western \textit{HST} field in the outer halo \citep{2007ApJ...666..903H, 2016ApJ...822...70L} resembles that of the resolved intra-cluster RGB stars in the Virgo Cluster core \citep{2007apj...656..756w}. Combined with the stellar kinematics discussed previously, we thus conclude that the population of PNe associated with the metal-poor exponential envelope trace the IGL of the Leo~I group. 

\subsection{Halo and IGL formation scenarios}
\citet{2016ApJ...822...70L} proposed a two-mode formation scenario for the metal-rich and metal-poor stellar populations in M105, in which the metal-rich \emph{inner halo} was formed in situ or through major mergers or relatively massive, and thus metal-rich progenitors. Later, the blue and metal-poor halo that we identify as the \emph{exponential envelope} in this work was assembled through dissipationless mergers and accretion. 

In addition to the inferences based on their metal-rich nature, the kinematics of stars in the inner halo point towards an in-situ origin or they may have been brought in through few massive and ancient mergers. The outwardly decreasing rotation and LOS velocity dispersion profiles are similar to those of in-situ stars in massive ETGs in cosmological hydrodynamical simulations such as IllustrisTNG \citep{2021A&A...647A..95P}. The formation of the metal-rich inner stellar halo through massive and ancient mergers is also observed in IllustrisTNG \citep{2021arXiv211013172Z}. The PN population properties, such as the lower $\alpha$-parameter value and shallower PNLF slope, are consistent with relatively massive and old parent stellar populations \citep{2006MNRAS.368..877B}. 

The blue and metal-poor exponential envelope instead is traced by a PN-rich population, whose high $\alpha$-parameter value is similar to that of Local Group dwarf irregular galaxies \citep[such as Leo I, and Sextans A and B;][]{2006MNRAS.368..877B}. The high velocity dispersion and moderate rotation of these PNe corroborate the late accretion scenario of \citet{2016ApJ...822...70L}.  Lastly, \citetalias{2020A&A...642A..46H} already noted that their luminosity estimate for the exponential envelope ($2.04\times 10^9\;L_\odot$) is similar to the luminosity of single ultra-faint galaxies (UFGs) in group and cluster environments \citep{2015ApJ...809L..21M}. \citet{2016ApJ...822...70L} postulate that UDGs are strong candidates to be responsible for the metal-poor stellar population in the exponential envelope of M105, as they have comparable metallicity distribution functions \citep[e.g.][]{2014ApJ...795L...6J}. Due to their low mass and density, UFGs can be easily stripped, and their debris can be deposited at large radii from the massive ETG \citep{2017mnras.464.2882a}, making them viable progenitors of the IGL stars. 

\subsection{The case of the ICL surrounding M49 in the Virgo Cluster}
\citet{2017A&A...603A.104H} showed that PN populations in the inner halo (within 60 kpc, corresponding to  $\approx 2.8 R_\mathrm{eff}$) and the outskirts of the ETG M49 in the massive ($10^{15}\;M_\odot$) Virgo Cluster have distinct spatial distributions and arise from stellar populations with different $\alpha$-parameters and with distinct PNLF slopes. Based on data from the PN.S, \citet{2018A&A...616A.123H} also showed that bright and faint PN populations have distinct kinematics, with the faint population tracing the transition to the ICL of the Virgo Subcluster B signalled by an increase of the LOS velocity dispersion as function of radius, reaching that of satellite galaxies orbiting the subcluster at large radii.

In the case of M49, \citet{2018A&A...616A.123H} had to rely on the combination of galaxy colours and PN dynamics to infer a link between the high $\alpha$-parameter measured in the transition region between the halo and ICL and an old, metal-poor underlying stellar population to explain the blue colours observed in the outer halo of M49 \citep{2013apj...764l..20m}. However, with studying the nearby galaxy M105, in the much less massive \citep[$6.4\times10^{11}\;M_\odot$;][]{1985ApJ...288L..33S} Leo~I group of galaxies, it is now possible to unambiguously link the presence of a PN population with a high $\alpha$-parameter \citepalias{2020A&A...642A..46H} to a co-spatial metal-poor stellar population forming part of the IGL of the Leo~I group.

\section{Summary and Conclusions}
\label{sec:summary}
In this paper, we have presented a new wide-field kinematic survey of PNe in the Leo~I group. We have discussed the sample selection and catalogue construction, as well as the overlap with previous photometric and kinematic surveys. The photometric and kinematic catalogues are included in Appendix~\ref{appendix} and \emph{will be made available in full through the CDS}. We have separated PNe into populations associated with the bulge and disk of the group galaxy NGC~3384, and with the S\'{e}rsic halo and exponential envelope of our main target, M105. 

We have identified three kinematically distinct populations of PNe in the halo of M105, whose properties we summarise in turn.
\begin{enumerate}
    \item The \emph{rotating core} within $1~R_\mathrm{eff}$ (2.7 kpc), characterised by a solar-metallicity stellar population that was likely formed {\it in situ}. 
    \item The \emph{inner halo}, from $1~R_\mathrm{eff}$ to $7.5~R_\mathrm{eff}$ (2.7 kpc to 20.25 kpc), made up by old intermediate-metallicity and metal-rich stars and characterised by low rotation and LOS velocity dispersion PNe with a low $\alpha$-parameter. The inner halo was either entirely formed {\it in situ}, or through notable contributions from massive and metal-rich merging events.
    \item The \emph{exponential envelope}, from $7.5~R_\mathrm{eff}$ (20.25 kpc) to the last data point at $16~R_\mathrm{eff}$ (43.2 kpc), containing a metal-poor and PN-rich stellar population with increasing rotation and constantly high LOS velocity dispersion. The exponential envelope was formed through dissipationless mergers and accretion of dwarf galaxies and very likely forms part of the extended IGL of the Leo~I group.
\end{enumerate}

Future work will focus on dynamical modelling of the PN subpopulations to determine the mass profile of M105 and its group environment. We will also carry out an in-depth comparison of the kinematic transitions identified in this work with new deep photometry. Lastly, this data set will also allow us to investigate changes of the PNLF bright cut-off for the kinematically distinct populations of PNe in the inner halo and exponential envelope.
\section*{Acknowledgements}
We are grateful to Nigel G. Douglas for his fundamental contribution to the Planetary Nebula Spectrograph. We greatly acknowledge the support and advice of the staff of the Isaac Newton Group on La Palma. We thank E. Iodice, R. Ragusa, and M. Spavone for the insightful discussions about the deep photometry of M105 from the VEGAS survey. AA was supported by Villum Investogator grant (proj.n. 16599) and a Villum Experiment grant (proj.n. 36225). JH acknowledges the hospitality of the Sub-department of Astrophysics of the University of Oxford. AJR was supported as a Research Corporation for Science Advancement Cottrell Scholar. CS is supported by a `Hintze Fellowship' at the Oxford Centre for Astrophysical Surveys, which is funded through generous support from the Hintze Family Charitable Foundation. 

This research made use of \textsc{astropy} \citep{2013A&A...558A..33A}, \textsc{astroplan} \citep{2018AJ....155..128M}, \textsc{astroquery} \citep{2019arXiv190104520G},  \textsc{astromatic-wrapper}, \textsc{corner} \citep{ForemanMackey2016}, \textsc{emcee} \citep{2013PASP..125..306F}, \textsc{lmfit} \citep{Newville_2014_11813}, \textsc{matplotlib} \citep{Hunter:2007}, and \textsc{numpy} \citep{2011arXiv1102.1523V}.
This research has made use of the SIMBAD database, operated at CDS, Strasbourg, France \citep{2000A&AS..143....9W}. This research has made use of the VizieR catalogue access tool, CDS,
Strasbourg, France.  The original description of the VizieR service was published in \citet{2000A&AS..143...23O}. This publication has made use of data products from the Two Micron All Sky Survey, which is a joint project of the University of Massachusetts and the Infrared Processing and Analysis Center/California Institute of Technology, funded by the National Aeronautics and Space Administration and the National Science Foundation. The Digitized Sky Surveys were produced at the Space Telescope Science Institute under U.S. Government grant NAG W-2166. The images of these surveys are based on photographic data obtained using the Oschin Schmidt Telescope on Palomar Mountain and the UK Schmidt Telescope. The plates were processed into the present compressed digital form with the permission of these institutions.
This research has made use of NASA's Astrophysics Data System.

\bibliographystyle{aa} 
\bibliography{literature}

\begin{appendix}
\section{Tabular data}
\label{appendix}
In this appendix we present the catalogues of PNe observed in the Leo~I group with Surprime-Cam at Subaru Telescope and the PN.S at the William Herschel Telescope. Table~\ref{tab:scam} provides the IDs, coordinates (J2000 with the \textit{2MASS} catalogue as astrometric reference), and $AB$ [\ion{O}{iii}] and $V$-band magnitudes of all PN candidates brighter than the limiting magnitude $m_\mathrm{5007,lim} = 28.1$, which were discovered based on Surprime-Cam observations. The survey objectives, data reduction, and PN candidate identification and validation are described in detail in the companion paper \citetalias{2020A&A...642A..46H} and briefly summarised in Sect.~\ref{ssec:photometry} of this work.

Table~\ref{tab:pns_final} provides the IDs, coordinates (J2000 with the \textit{2MASS} catalogue as astrometric reference), $AB$ [\ion{O}{iii}] magnitudes, LOS velocities and corresponding errors, as well as the membership probability to be associated with M105 (see Sect.~\ref{sec:decomposition}). Two concatenated IDs denote PNe that were observed in two fields, and objects with IDs starting with M105-C and N3384 were observed first by \citet{2007apj...664..257d} and \citet{2013a&a...549a.115c} respectively.  The catalogue only contains PNe with velocities within $3\sigma$ about the robust mean (see Sect.~\ref{ssec:kincat}). \textit{Both catalogues will be made available in full through the CDS.}

\begin{table*}
\centering
\caption{IDs, coordinates, magnitudes, and velocities of PNe candidates from Surprime-Cam photometry \citepalias{2020A&A...642A..46H}. Only objects brighter than the limiting magnitude are included. \textit{The full table will be made available through the CDS.}}
\label{tab:scam}
\begin{tabular}{lllll}
\hline
\hline
ID & RA & dec & $m_{[\ion{o}{iii}], AB}$ & $V_{AB}$ \T \\
 & [hh:mm:ss] (J2000)  & [$\degr$:\arcmin:\arcsec] (J2000) & [$\mathrm{mag}$] & [$\mathrm{mag}$] \B \\
\hline
M105-SCAM-001 & 10:47:21.6842 & 12:23:27.1824 & 25.3 & 27.3 \T \\
M105-SCAM-002 & 10:47:40.7357 & 12:23:33.2999 & 25.0 & 26.4 \\
M105-SCAM-003 & 10:47:29.4416 & 12:23:45.1158 & 25.1 & 27.3 \\
$\vdots$ & $\vdots$ & $\vdots$ & $\vdots$ & $\vdots$ \\
M105-SCAM-226 & 10:47:17.8139 & 12:48:02.209 & 25.5 & 28.7 \\
\hline
\hline
\end{tabular}
\end{table*}

\begin{table*}
\centering
\caption{IDs, coordinates, magnitudes, velocities and membership probabilities of PNe in the Leo~I galaxies M105 and NGC~3384. Objects with IDs starting with M105-C and N3384 were observed first by \citet{2007apj...664..257d} and \citet{2013a&a...549a.115c} respectively. Two concatenated IDs denote PNe that were observed in two fields. The catalogue only contains PNe with velocities within $3\sigma$ about the robust mean (see Sect.~\ref{ssec:kincat}).  \textit{The full table will be made available through the CDS.}}
\label{tab:pns_final}
\begin{tabular}{llllrll}
\hline
\hline
ID & RA & dec & $m_{[\ion{o}{iii}], AB}$ & $v_\mathrm{los}$ & $\Delta v_\mathrm{los}$ & $f_\mathrm{M105}$ \T \\
 & [hh:mm:ss] (J2000)  & [$\degr$:\arcmin:\arcsec] (J2000) & $[\mathrm{mag}]$ & $[\mathrm{km\,s^{-1}}]$ & $[\mathrm{km\,s^{-1}}]$ & \B \\
\hline
M105-C001\_M105-W001 & 10:47:31.1951 & 12:39:20.1428 & 24.9 & 1002.0 & 19.1 & 0.99631 \T\\
M105-C002 & 10:47:32.2051 & 12:32:09.7428 & 24.3 & 909.0 & 20.0 & 0.99354 \\
M105-C003 & 10:47:33.5851 & 12:34:08.5428 & 25.4 & 892.0 & 20.0 & 0.99072 \\
$\vdots$ & $\vdots$ & $\vdots$ & $\vdots$ & $\vdots$ & $\vdots$ & $\vdots$\\
N3384-C093 & 10:48:33.2169 & 12:36:27.0622 & 23.8 & 629.0 & 20.0 & 0.01520 \\
\hline
\hline
\end{tabular}
\end{table*}

\end{appendix}
\end{document}